\documentclass[useAMS,usenatbib]{mn2e}

\usepackage{graphics}
\usepackage{epsfig}
\usepackage{latexsym}
\usepackage{graphics}
\usepackage{graphicx}
\usepackage{epstopdf}
\usepackage{url}

\newcommand{\be}{\begin{equation}}
\newcommand{\ee}{\end{equation}}
\newcommand{\ba}{\begin{eqnarray}}
\newcommand{\ea}{\end{eqnarray}}
\newcommand{\bc}{\begin{center}}
\newcommand{\ec}{\end{center}}

\title{Building up the spectrum of cosmic-rays in star-forming regions}

\author[Torres, Cillis, Lacki, \& Rephaeli]
{Diego F. Torres$^{1,2}$, Anal\'ia Cillis$^{3}$, Brian Lacki$^{4}$, \& Yoel Rephaeli$^{5,6}$\\
$^1$Instituci\'o Catalana de Recerca i Estudis Avan\c{c}ats (ICREA) Barcelona, Spain \\ 
$^2$Institut de Ci\`encies de l'Espai (IEEC-CSIC), Campus UAB,  Torre C5, 2a planta, 08193 Barcelona, Spain \\
$^3$Instituto de Astronom\'ia y F\'isica del Espacio (CONICET-UBA), CC67, Suc. 28, 1428 Buenos Aires, Argentina \\
$^4$Jansky Fellow, Institute for Advanced Study, Einstein Drive, Princeton, New Jersey, 08540 USA\\
$^5$Raymond and Beverly Sackler School of Physics and Astronomy, Tel Aviv University, Tel Aviv, 69978, Israel\\
$^6$Center for Astrophysics and Space Sciences, University of California, San Diego, La Jolla, CA 92093, USA
}

\begin{document}

\date{}

\pagerange{\pageref{firstpage}--\pageref{lastpage}} \pubyear{2012}

\maketitle

\label{firstpage}

\begin{abstract}

The common approach to compute the cosmic-ray distribution in an starburst galaxy or region is equivalent to assume that at any point within that environment, there is an accelerator inputing cosmic rays at a reduced rate. This rate should be compatible with the overall volume-average injection, given by the total number of accelerators that were active during the starburst age. These assumptions seem reasonable, especially under the supposition of an homogeneous and isotropic distribution of accelerators. However, in this approach the temporal evolution of the superposed spectrum is not explicitly derived; rather, it is essentially assumed ab-initio. Here, we test the validity of this approach by following the temporal evolution and spatial distribution of the superposed cosmic-ray spectrum and compare our results with those from theoretical models that treat the starburst region as a single source. { In the calorimetric limit (with no cosmic-ray advection), homogeneity is reached (typically within 20\%) across most of the starburst region. However, values of center-to-edge intensity ratios can amount to a factor of several. Differences between the common homogeneous assumption for the cosmic-ray distribution and our models are larger in the case of two-zone geometries, such as a central nucleus with a surrounding disc. We have also found that the decay of the cosmic-ray density following the duration of the starburst process is slow, and even approximately 1 Myr after the burst ends (for a gas density of 35 cm$^{-3}$) it may still be within an order of magnitude of its peak value.}
Based on our simulations, it seems that the detection of a relatively hard spectrum up to the highest gamma-ray energies from nearby starburst galaxies favors a relatively small diffusion coefficient (i.e., long diffusion time) in the region where most of the emission originates.

\end{abstract}


\section{Introduction}

From radio to TeV gamma-rays, starbursts shine in the sky.  Their large star formation rates power high injection rates of cosmic rays in the interstellar media of these galaxies, which emit nonthermal radiation. The large CR energy contents in starbursts are of interest for cosmic ray feedback (Socrates et al. 2008), ionization (Papadopoulos 2010), the gamma-ray background (e.g., Thompson, Quataert \& Waxman 2007; Makiya et al. 2011), and  for determining whether they are neutrino sources (e.g., Loeb \& Waxman 2006).  Indeed, the large cosmic content has been supported by the recent detections of M82 and NGC 253 by GeV and TeV telescopes (Abdo et al. 2010, Acciari et al. 2009, Acero et al. 2009).  In anticipation and as spinoff of these detections, there has recently been a spate of modelling of the cosmic ray spectrum in starburst galaxies.

In considering the cosmic ray spectrum in star forming regions, models generally start by writing the
diffusion-loss equation   (see, e.g.,
Ginzburg \& Syrovatskii 1964) 
\ba
&& - D \bigtriangledown ^2 N(E)+\frac{N(E)}{\tau(E)} - \frac{d}{dE}
\left[ b(E) N(E) \right] - Q(E) = \nonumber \\  && \hspace{5cm}
 - \frac{\partial N(E)}{\partial
t} ,
\label{DL} 
\ea 
where $N(E)$ is the distribution of particles
with energies in the range $E$ and $E+dE$ per unit volume, and
$b(E)=-\left( {dE}/{dt} \right)$ is the rate of loss of energy.
In this equation, $D$ is the scalar diffusion
coefficient, $Q(E)$ represents the source term appropriate to the
production of particles with energy $E$, and $\tau(E)$ is the residence time
(beyond which particles are removed from the phase space).
It is usually assumed that the starburst is in a steady state,  $ {\partial N(E)}/{\partial t}
=0,$ and, under the assumption of a homogeneous distribution of
sources, that the spatial dependence can be ignored, so
that $ D \bigtriangledown ^2 N(E) =0$. These assumptions were adopted by essentially all detailed models of
starburst regions published in the literature so far
(e.g., see Paglione et al. 1996, Blom et al. 1999, 
Torres 2004, 
Domingo-Santamaria \& Torres 2005, 
de Cea del Pozo et al. 2009, Lacki et al. 2010, and references therein).   
Persic et al. 2008, and Rephaeli et al. 2010 solve the 
diffusion-convection equation in 3D, but  do not discuss the detailed spatial distribution of cosmic-rays, just the average spectrum over the starburst and the surrounding disk. This was also the case of some of the formerly quoted papers. On the other hand, the GALPROP models (see Strong \& Moskalenko 1998) for the propagation of cosmic-rays in our Galaxy do not consider the formation of the cosmic-ray spectrum from the individual contributors, in a time-dependent manner.  

In the previously mentioned models, and in regards of Eq. (\ref{DL}),
the  proton injection is assumed to
be a power-law with index $p$, for instance, 
\be
Q_{\rm inj}(E_{\rm p,\, kin}) = K  {E_{\rm p,\, kin}}^{-p},
\ee
where $K$ is a normalization constant and units are
such that $[Q]$= {\rm GeV}$^{-1}$ {\rm cm}$^{-3}$  {\rm s}$^{-1}$. The proton injection is assumed to be uniform across the starburst.  The power law spectrum follows from the results of Fermi first order acceleration assumed for individual accelerators (e.g., Bell 1978, see also Ellison et al. 2004).
The normalization is obtained from the total power transferred
by supernovae into the CRs kinetic energy within a given volume 
\ba
\int_{E_{\rm p,\,kin,\, min}}^{E_{\rm p,\, kin,\, max}} Q_{\rm
inj}(E_{\rm p,\, kin}) E_{\rm p,\, kin} dE_{\rm p,\, kin} \approx \nonumber \\
-K
{E_{\rm p,\,kin,\,min}^{-p+2}}/(-p+2) \equiv 
\nonumber \\
{ \sum_i \eta_i {\cal P} {\cal R}_i }/ {V},
\label{inj}
\ea
where, for simplicity, $p\neq 2$ was assumed, and we used the fact that
$E_{\rm p,\,kin,\,min} \ll E_{\rm p,\,kin,max}$ in the second
equality. The last part of Eq. (\ref{inj}) is simply the total power using
${\cal R}_i$ ($\sum_i {\cal R}_i={\cal R}$) 
as the rate of supernova explosions in the star forming region of volume $V$ that release 
a fraction $\eta_i$ of its power ${\cal P}$ into relativistic cosmic rays.
We note that nonlinear theories of CR acceleration predict deviations from the power law injection spectrum, especially at low energies and near the high energy cutoff (e.g., Berezhko \& Ellison 1999; Ellison et al. 2000).  In addition, Ptuskin \& Zirakashvili 2005 argue that at any given time, only particles near an evolving cutoff energy actually escape the supernova remnant, and only the time-averaged injection spectrum is a power law (see also Fujita et al. 2009; 2010).  Thus, the spectrum and time evolution of CR accelerators may be more complicated than we consider here.

The injection spectrum of the primary particles ($Q(E)$, from which one derives the distribution $N(E)$ and then e.g., the gamma-ray emission) is assumed to be directly normalized by the total supernovae explosions in the galaxy, and to have a power-law form with fixed index, irrespective of position within the starburst. The common approach is then equivalent to assuming that  at any point of the starburst region, there is a single accelerator
inputing cosmic rays at a reduced rate, one that is compatible with the overall volume-average given by the total number of accelerators.
How realistic are all these assumptions?

There are direct observational hints that the luminosity in gamma-rays is correlated with the star formation rate (from now on, SFR)  or supernova frequency. The correlation may be slightly non-linear, following 
a power-law with a hard index, $L_\gamma \sim$ SFR$^b$, with $b \sim 1 - 1.4$ (Abdo et al. 2010, 2011). 
Due to the current scarcity of starbursts measured in gamma-rays, this is, however, not yet proven.
Strictly speaking,
this relation is in fact impossible to predict in the models presented in the literature so far, since
the SFR goes into the normalization of the primary cosmic ray spectrum, which is assumed. Then, 
it is natural to expect that to first order, the same correlation is maintained in the hadronic gamma-ray luminosity generated by the former population.
Plots such as  $L_\gamma$ vs.  SFR (or similarly, enhancement  of cosmic-rays vs SFR) 
have not been derived (to our knowledge)  from first principles yet. 

We also note that the problem of calorimetric (essentially, full conversion of the proton kinetic energy into secondary particles like gamma-rays) or non-calorimetric behavior
also depends on the validity of the assumption of the primary spectrum, emphasizing that an understanding on how it is built up from individual contributors could be useful. 

The aim of this work is then 
to assess some of the formerly-commented common assumptions made in the description of the injected cosmic-ray
spectrum of starbursts regions,
by accounting for the individual contributions of accelerators residing in the starburst volume over the starburst age.

\section{Numerical approach and results}

As stated, instead of considering a total, volume-averaged, cosmic-ray injection, 
we shall obtain the individual contribution of each accelerator to the cosmic-ray intensity at a given time.
The cosmic ray intensity contribution of each one of the accelerators will thus be given by 
\be
   J_p(E,\;r,\;t)= \frac {c}  {4\pi} \;  f,
\ee
where $f(E,\;r,\;t)$ is the distribution function of protons at
time $t$ and distance $r$ from the source which
satisfies the radial-temporal-energy dependent diffusion equation 
\ba
   && ({\partial f}/{\partial t})=({D(E)}/{r^2}) ({\partial}/{\partial
   r}) r^2 ({\partial f}/{\partial r}) +\nonumber \\  
   && \hspace{4cm} ({\partial}/{\partial
   E}) \, (Pf)+Q
   \label{DL2}
\ea
(we choose spherical coordinates, with $R$ being  the radial
distance from a given accelerator, and $P$ representing the energy losses).
We should note the differences between the simplified Eq. (\ref{DL}), for the whole starburst, and Eq. (\ref{DL2}). The latter is written
for a {\it single accelerator} and thus, $Q$ is the injection of cosmic-rays of that accelerator only and  is obviously more constrained.
Note too that  Eq. (\ref{DL2}) 
refers to  an $E-$, $R-$, and $t-$dependent problem, whereas  after the simplifications over 
Eq. (\ref{DL}) that we mentioned above, it is
homogeneous and in steady state, and thus only $E-$dependent. 
But we note a cavaet too: in order to work with analytically tractable solutions of Eq. (\ref{DL2}), see below, we ignored the advective escape term (i.e., $N(E)/\tau(E)$); thus, we do not consider here outflows that may remove cosmic rays from the starburst regions.  Cosmic rays are transported here by diffusion, which is appropriate in the calorimetric limit only (see, e.g., Loeb \& Waxman 2006, Thompson et al. 2007).. 

\begin{figure}
\center
\includegraphics[width=0.4 \textwidth]{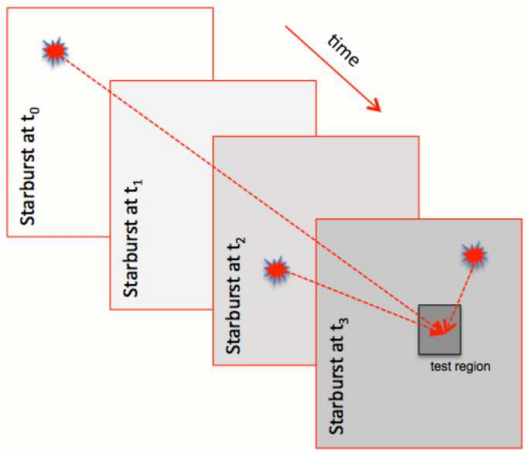}
\caption{Concept for the computation of individual contributions of the cosmic-ray density in star forming regions.} 
\label{concept}
\end{figure}

We assume that the proton energy loss is due to
nuclear interactions.
The nuclear loss rate is $P_{\rm nuc} = E/\tau_{pp}$, with $\tau_{pp}=(n_p\, c \, \kappa \, \sigma_{pp} ) ^{-1}$ is the timescale for the corresponding nuclear loss, $\kappa \sim 0.45$ is the inelasticity of the interaction, and $\sigma_{pp} \sim 33$ mb is the cross section (e.g., Stecker 1971, Gaisser 1990); $\sigma_{pp}$ is essentially constant for the energies of interest above GeV. Thus, following e.g., Aharonian \& Atoyan (1996), 
$\tau_{pp} \sim 7 \times 10^7 (n/{\rm cm}^{-3})^{-1}$ yr.
The solution to the diffusion equation for an arbitrary energy loss term, diffusion coefficient, and impulsive  injection spectrum $f_{\rm inj}(E)$, such that  $Q(E,r,t) = N_0 f_{\rm inj}(E) \delta{\bar r} \delta(t)$, for 
the particular case in which $D(E)\propto E^\delta$ and $f_{\rm inj}\propto E^{-\alpha}$,   is 
(see Aharonian \& Atoyan 1996 and references therein) 
\ba
  && f(E, r,t) \sim ({N_0 E^{-\alpha}}/{\pi^{3/2} R_{\rm dif}^3}) 
  \nonumber \\  
   && \hspace{1cm} \times
   \exp \left[ { - {(\alpha-1)t}/{\tau_{pp} }- ({R}/{R_{\rm dif}})^2} \right],
  \label{sol}
\ea
 where 
\be 
R_{\rm dif} = 2 \sqrt{ D(E) t \frac{\exp(t \delta / \tau_{pp})-1}{(t \delta / \tau_{pp})}}
\label{rdii}
\ee
 stands for the radius of the sphere up to which the particles of energy $E$ have time to propagate after their injection.
 In case of continuous injection of accelerated particles, given by $Q(E, \;
t)=Q_0 E^{-\alpha} {\cal T}(t)$, the previous solution needs to be convolved with the function ${\cal T}(t-t')$ in the time interval $0 \leq t' \leq t$. 
If the source is described by a Heavside function,  ${\cal T}(t)=\Theta (t)$ the solution is 
(Atoyan et al. 1995) 
\ba
&& f(E,\;r,\;t)=({Q_0 E^{-\alpha}}/{4\pi D(E) r}) (
{2}/{\sqrt{\pi}})\nonumber \\  
   &&
   \hspace{3cm} 
\times \int^{\infty}_{r/R_{\rm diff}} e^{-x^2}
dx.
\ea
All previous equations in this section are thus valid for a single accelerator, e.g., a single supernova remnant or pulsar wind nebula, injecting either impulsively or continuously cosmic rays into the starburst.
We now consider that a starburst is  a collection of such accelerators, which are assumed to 
appear in the starburst volume 
at a given rate, e.g.,  equal to the SN rate (which is in turn related to the SFR), and which 
could be constant (as is usually assumed) or not (e.g., the starburst phenomenon could have ended, or it could be a short burst of star formation rather than a continuous one).
Thus, one can consider the evolution in time of the injection spectrum of the whole starburst 
by computing each accelerator's contribution using the solution to Eq. (\ref{DL2}), and solving it as many times
as there are injectors in the volume. 

This concept is depicted in Figure \ref{concept}. Each plane represents a given past 
time in the evolution of the starburst (the current time is $t_3$ in that figure). Each test region of the starburst today receives
the cosmic-ray contribution of each of the accelerators injecting cosmic-rays in its past, for which we individually consider
the propagation from the time of its injection up to the test region. Summing up all contributions, we build up the spectrum at that point.
Comparing different test points we can check for homogeneity, and study, for instance, how many and which are the accelerators 
contributing mostly to the total spectrum. 
One consequence of such an approach to compute the total spectrum is a direct relation between the accelerator appearance rate and
the cosmic-ray intensity in a given position. By slicing in time and space bins from that test-position, the number of accelerators
in each bin will, under a homogeneous and isotropic distribution, be directly proportional to the accelerator appearance rate. (Each of the panels
in Figure \ref{concept} will have as many accelerators as determined by the appearance rate assumed.)
Such proportionality would be apparent in the cosmic-ray intensity, since, for sufficiently small time and space bins, 
the differences among the contributions of accelerators that were born in that bin will be negligible; this would lead to a linear relationship between cosmic-ray intensity and accelerators appearance rate. We have checked that this is indeed the case in our numerical simulations.

\subsection{Cosmic-ray injection from a single accelerator}

\begin{figure}
\center
\includegraphics[width=8.0cm]{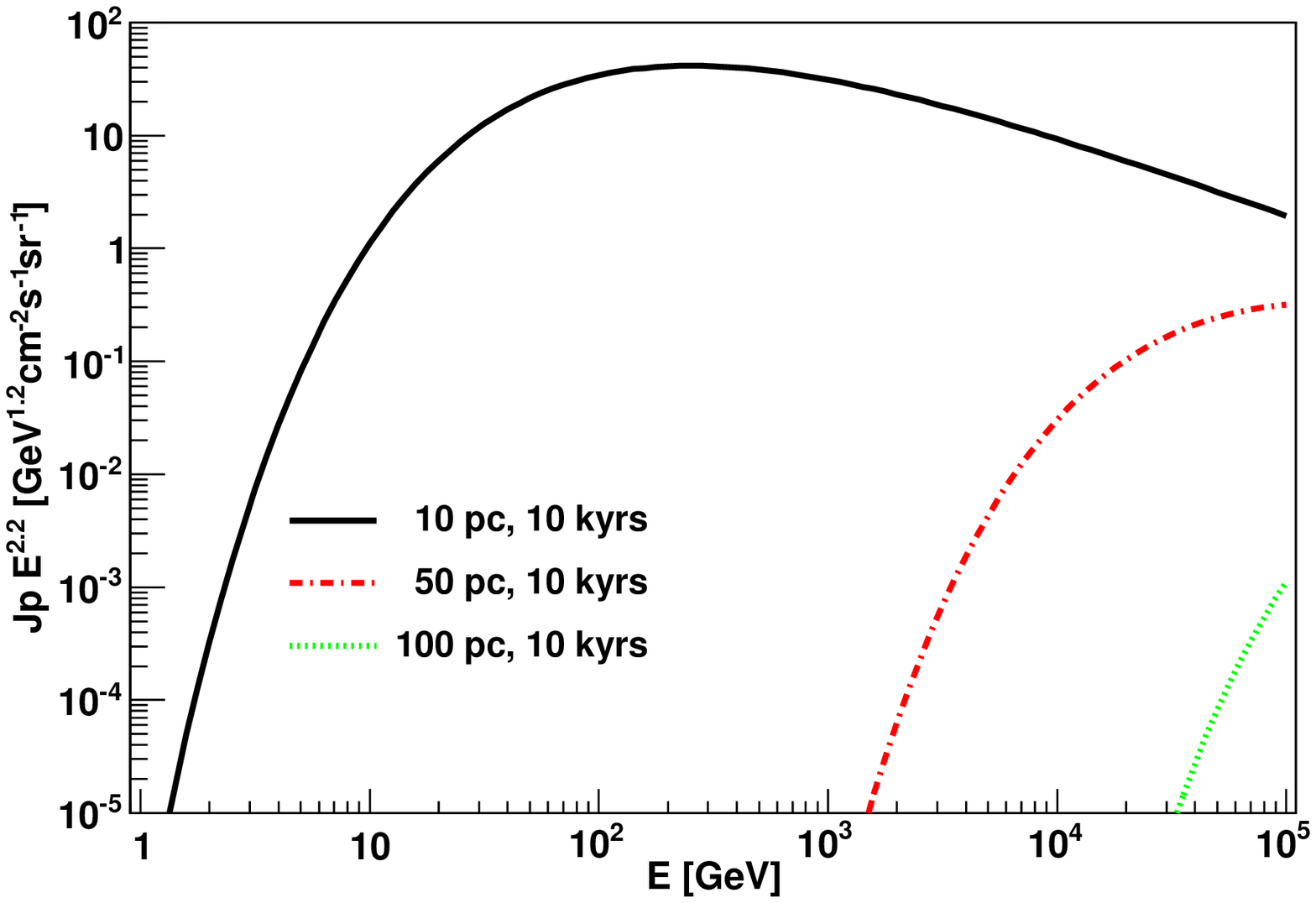}
\includegraphics[width=8.0cm]{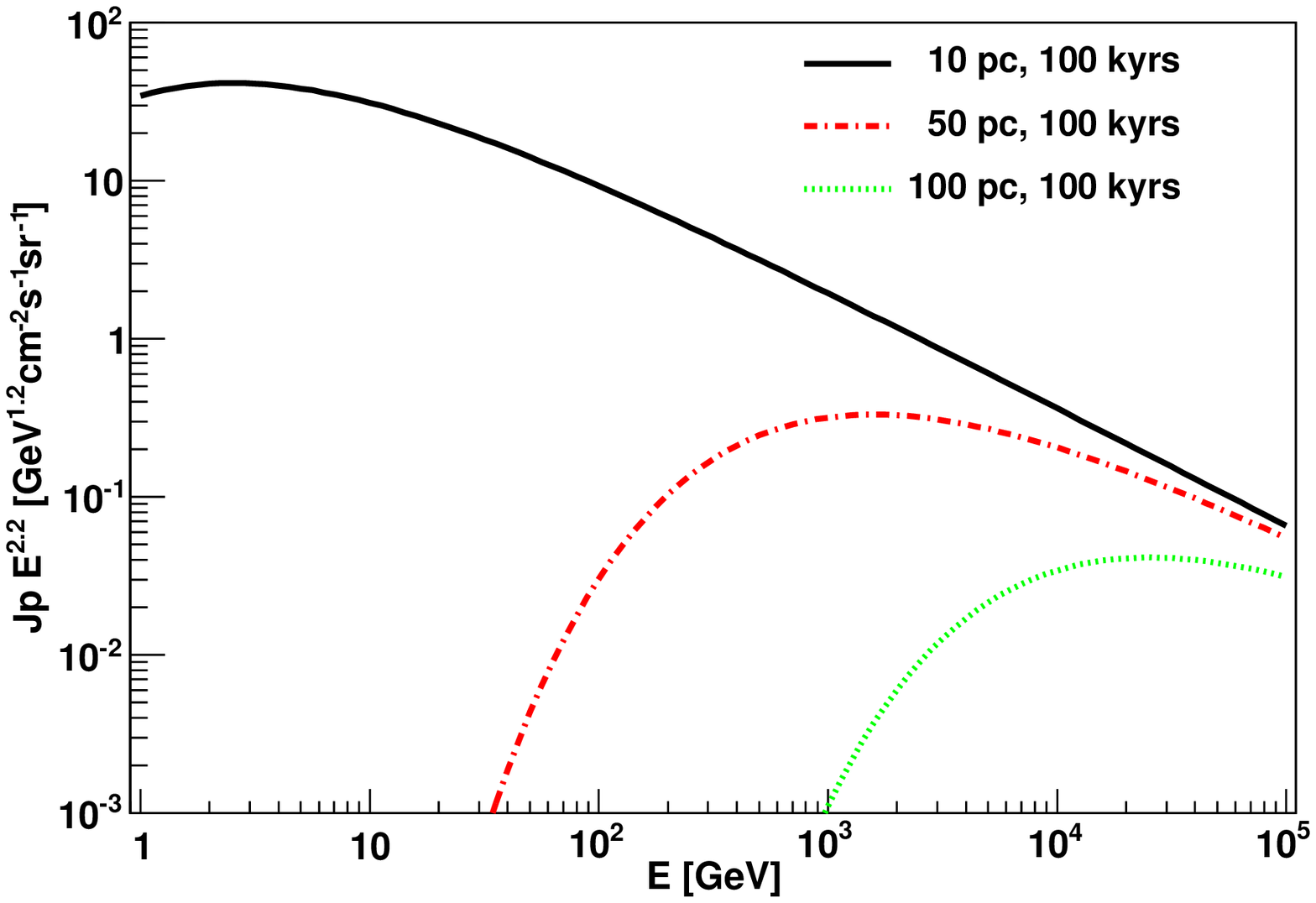}
\includegraphics[width=8.0cm]{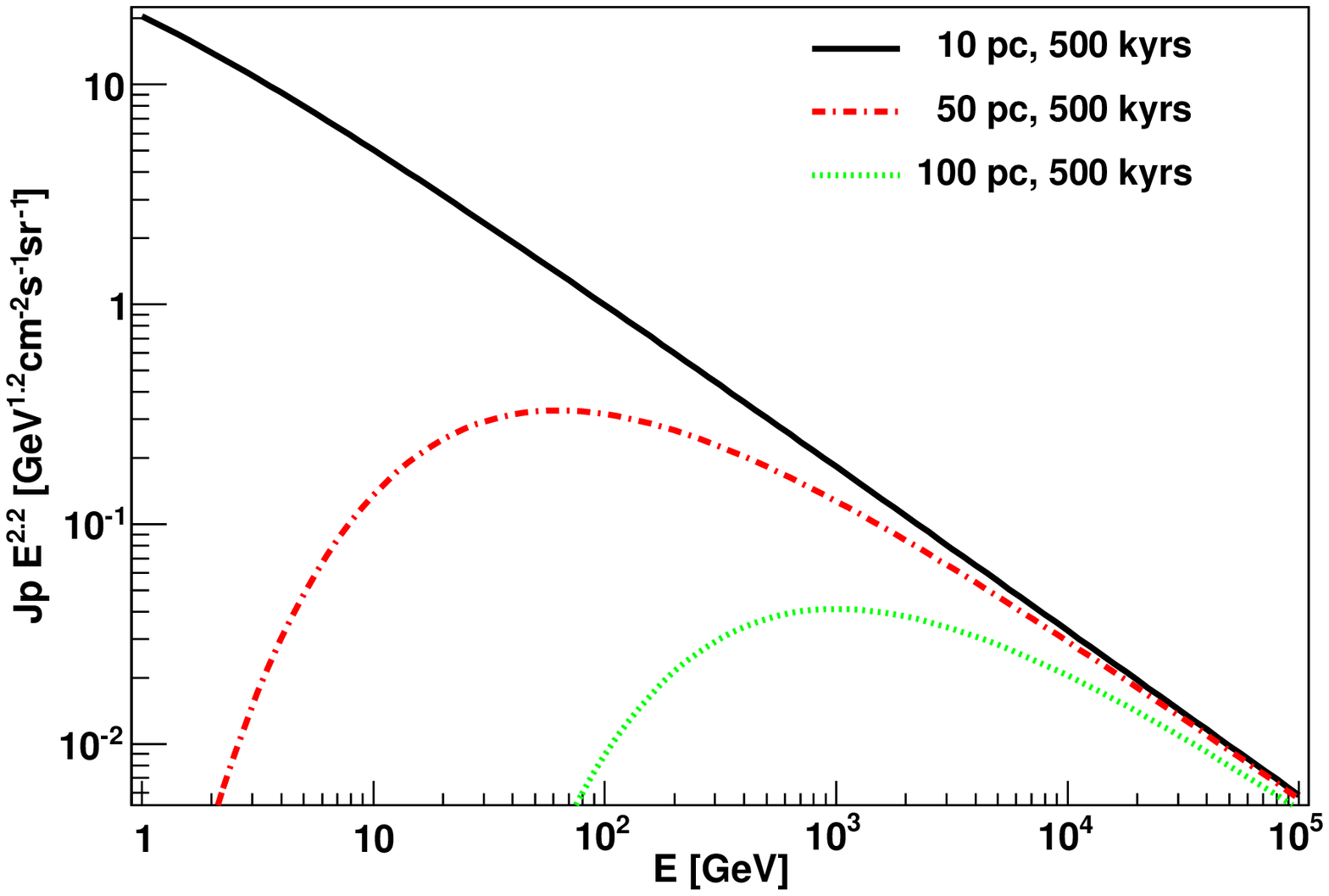}
\caption{Cosmic-ray intensity contribution for injection at different times in the starburst history, and at different distances from the test point. Times considered are, from top to bottom, 
$10^4$, 10$^5$ and $5 \times 10^5$ years. 
}
\label{single}
\end{figure}

\begin{figure}
\center
\includegraphics[width=8.0cm]{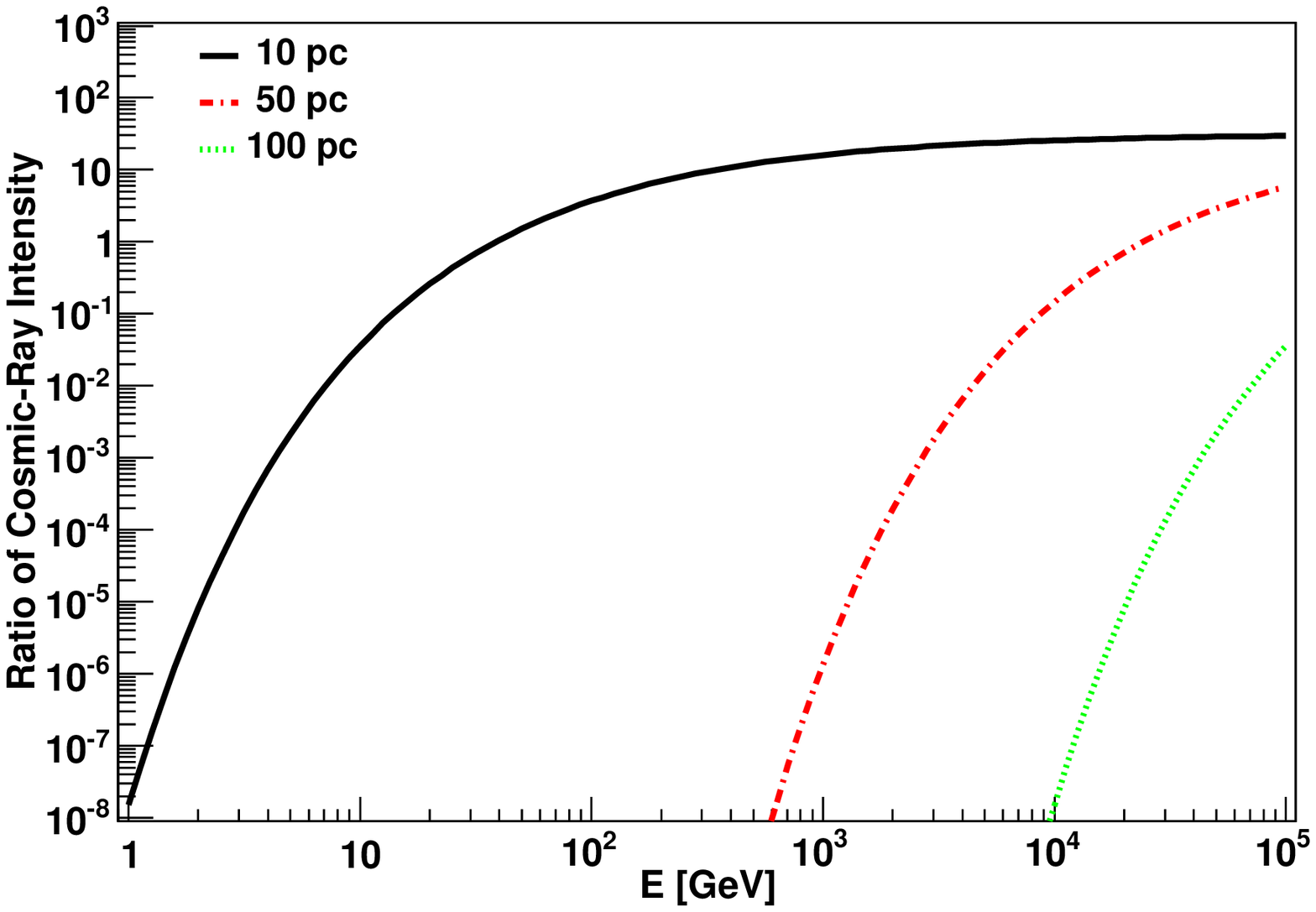}
\includegraphics[width=8.0cm]{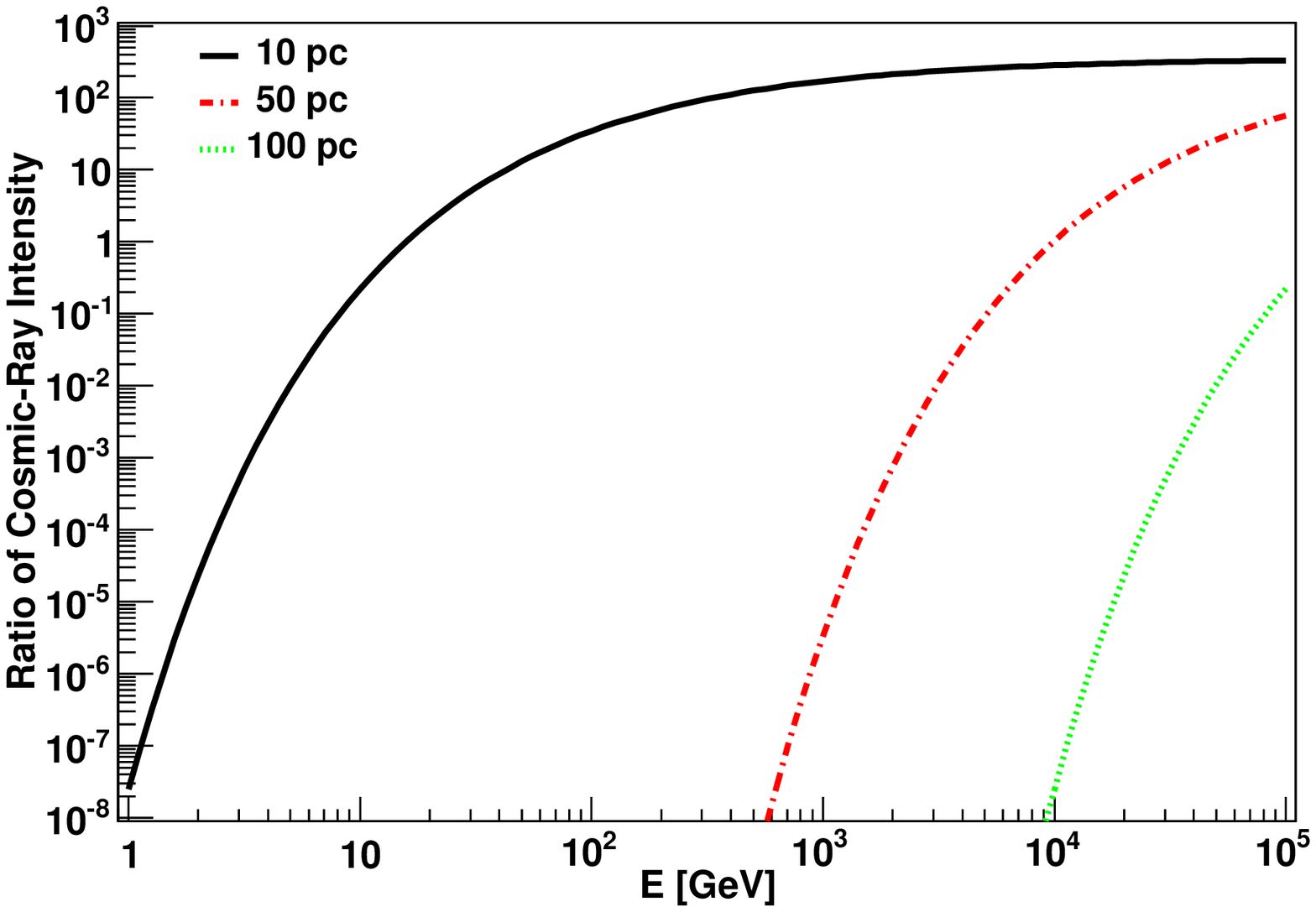}
\caption{
Ratio of the cosmic-ray intensity as a function of energy between a young and an old accelerator; i.e. one that injected particles at the earlier times of $10^4$ and $10^5$ years (top), and $10^4$ and $5 \times 10^5$ years (bottom), for the three selected distances to the test point. 
}
\label{ratio-single}
\end{figure}

As an example of the results, we consider first the injection of a single accelerator. We show in Figure \ref{single} the cosmic-ray intensity contribution for an injection that occurred at different times in the starburst history, and at different distances from the test point. We consider here (as we do below) that the diffusion coefficient at 10 GeV  is $10^{26}$ cm$^2$ s$^{-1}$ and $\delta=0.5$ in a medium of density $n=35$ cm$^{-3}$; we have also explored other parameter values.
A low value of diffusion coefficient (slow diffusion) is expected in dense molecular regions, as an starburst (see e.g., the cases of dense molecular clouds in our Galaxy in the works by Ormes et al. 1988, Gabici 2010, or Torres et al. 2010, and other references therein).
With the selected value of interstellar particle density,  the total molecular mass in a typical starburst sphere of 300 pc is 
$\sim 10^8$ M$_\odot$. 
Times considered are
$10^4$, 10$^5$ and $5 \times 10^5$ years and distances considered are 10, 50 and 100 pc from the test point. 

Figure \ref{single} 
can be understood assuming that 
$t \delta \ll \tau_{pp}$, and thus, from Eq. (\ref{rdii}) and using the adopted parameters, 
one gets the diffusion radius as 
$R_{\rm dif} \sim 12\ {\rm pc} \ (E / 10\ {\rm GeV})^{1/4} (t / 10^5\ {\rm yr})^{1/2}$.
We see that only high-energy particles diffuse away fast enough to reach distant regions. We also note that the ages of the accelerators we (purposefully) choose are reflected in low individual contributions to the cosmic-ray intensity. They have, however, a varying $E-$dependent importance of which injectors are the dominant contributors. 

{ This effect is more clearly seen in Figure 3, in which we plot the ratio of the cosmic-ray intensity as a function of energy, between an accelerator that
injected particles 10$^4$ and 10$^5$ years ago (top), and between 10$^4$ and $5
\times 10^5$ years ago (bottom). 
At low energies and distances, the contribution of the older accelerator dominates that of the younger one. At 10 GeV, for instance, the diffusion radius of the 10$^4$ year old accelerator is $\sim 4$ pc; whereas it is $\sim 12$ pc for the 10$^5$ year old one.  At 10 pc then, the older accelerator dominates. At higher energies, the diffusion radius of the 10$^4$ year old accelerator is larger than the test point considered and starts dominating.} These results are consistent with the study by Aharonian \& Atoyan (1996), and are just put here in the context of injection ages that are much larger than those typically considered for a single source (e.g., when studying the interaction of such cosmic-rays with the surrounding molecular gas, like in mid-age SNRs).

\subsection{Homogeneous starbursts at constant explosion rate}

Taking into account the individual contributions, we start by supposing that the starburst is a sphere (e.g., with a radius of 300 pc)
and has experienced a constant supernova explosion rate (e.g., of 0.1 SN year$^{-1}$) during the recent history (e.g., during the last $1\times 10^6$ years). With the previous values, we have a volume-average number of explosions (generically called accelerators or injectors) 
equal to $\sim 9\times 10^{-4}$ pc$^{-3}$.
In accord with the approach described above, the total cosmic ray distribution is
\begin{equation}
J_{p}=\sum_{n=1}^N\frac{c}{4\pi}f_{n}(E,d_{n},t_{n}),
\end{equation}
where $f_n$ is the individual cosmic-ray contribution. In this example, then, we numerically consider $10^5$ individual solutions
to Eq. (\ref{DL2}).
For this purpose we developed a numerical scheme that solves the diffusion-loss equation for each injector,
under the assumption that the cosmic-ray source is impulsive if the time  past since its appearance 
is larger than the adopted regime break of 10$^4$ years and continuous otherwise, and sums them all up. 
We randomize via  Monte Carlo the position and time of appearance of each accelerator (with the specified mean explosion rate), as well as the individual parameters, in particular, the energy input is randomized between $(0.1-1) \times 10^{50}$ erg for an impulsive case, and  between $(0.1-1) \times 10^{37}$ erg s$^{-1}$ for a continuous case. 
We again assume that the diffusion coefficient at 10 GeV  is $10^{26}$ cm$^2$ s$^{-1}$ and $\delta=0.5$ in a medium of density $n=35$ cm$^{-3}$. 
Each source spectrum was taken to have a slope of 2.2, but we also considered the case where the slope was randomly selected in an interval around this value.

\begin{figure}
\center
\includegraphics[width=8.0cm]{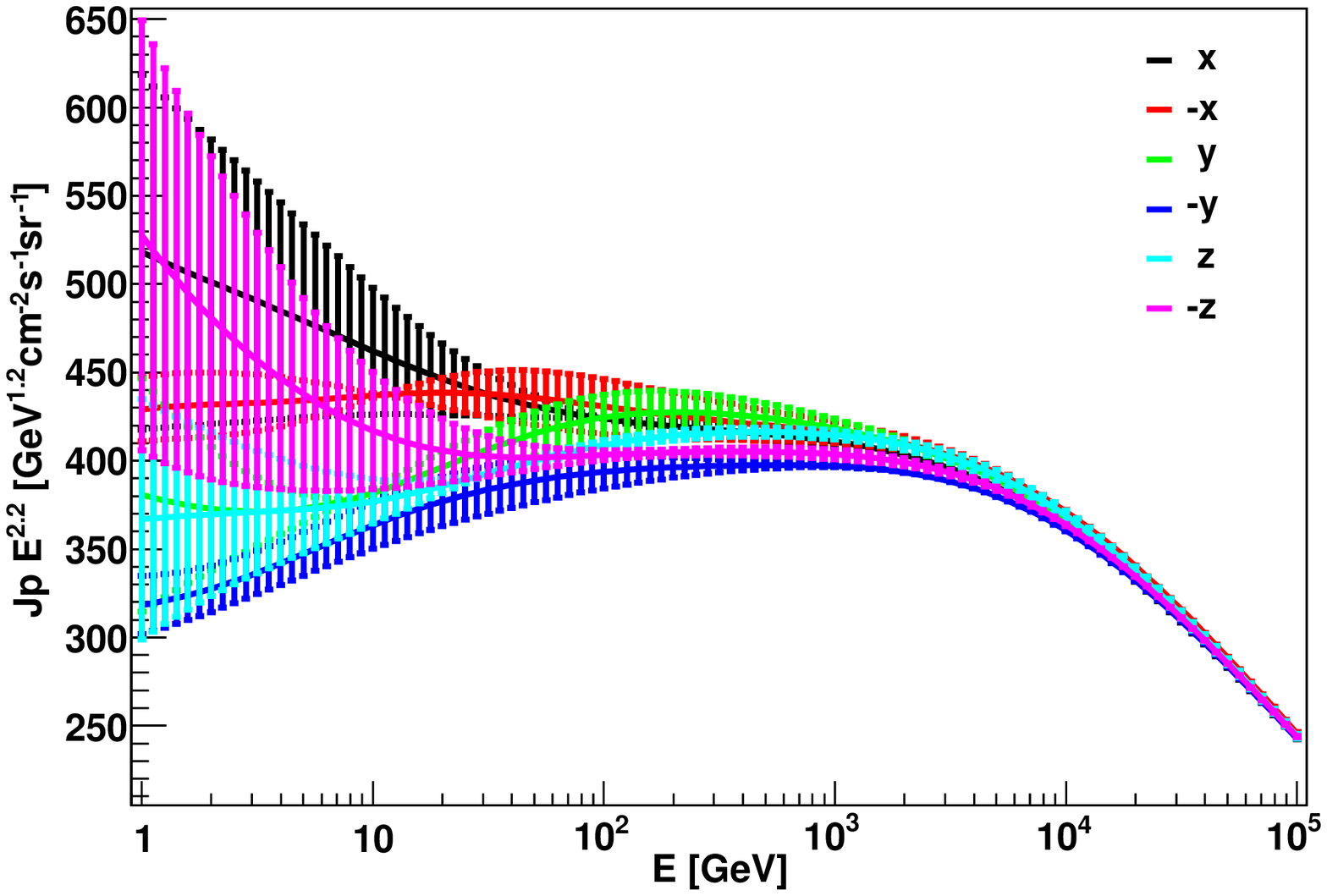}
\includegraphics[width=8.0cm]{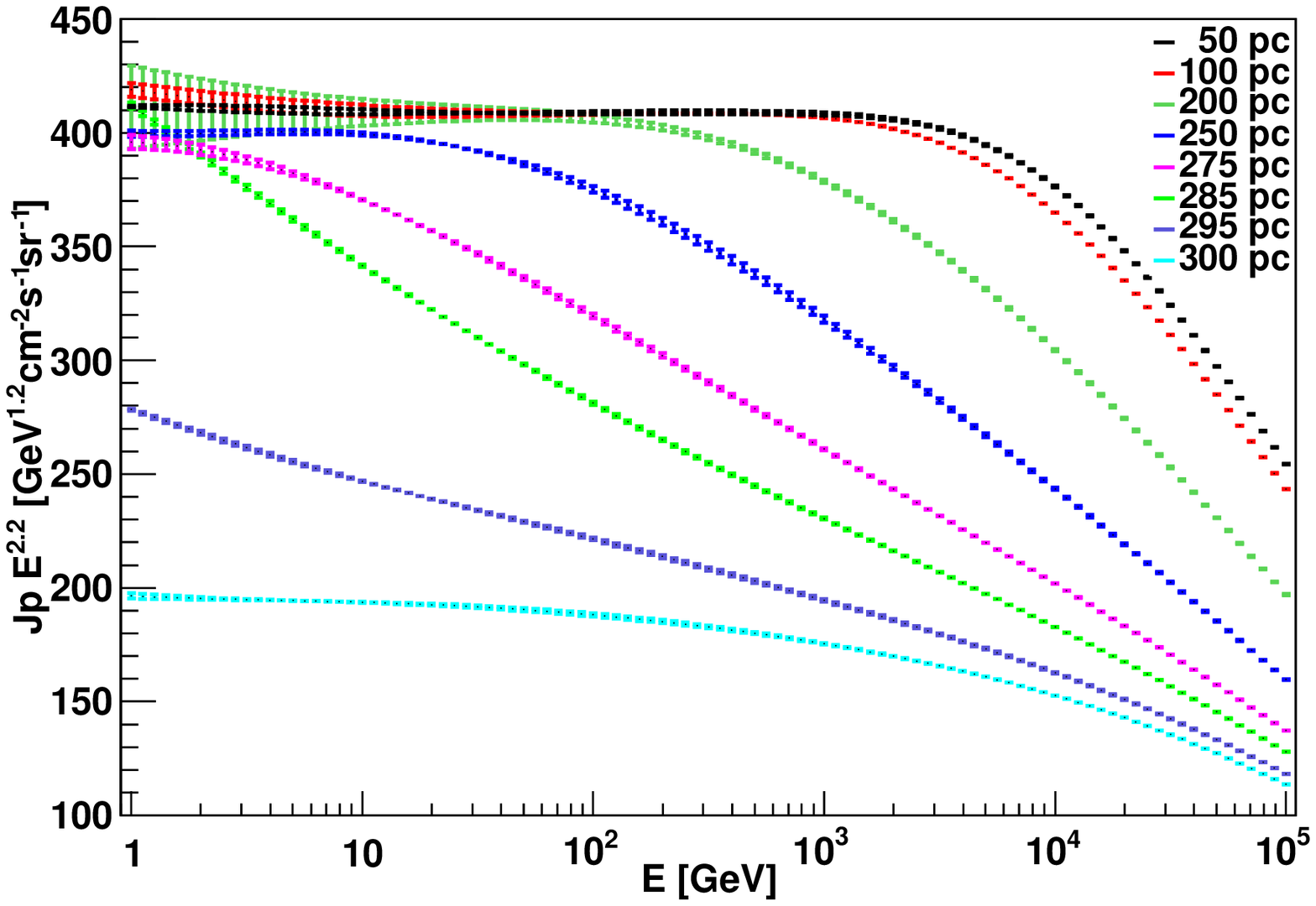}
\caption{
Top:
Cosmic-ray intensity  for antipodal positions in azimuth and zenith angle, all at a distance of 100 pc from the center of the starburst. Errors result from a small number of simulations (6 realizations only), and are shown merely for illustration. The larger the number of simulations, the closer all curves are to one another. For the sake of  clarity of presentation, the (ordinate) flux was multiplied by the factor $E^{2.2}$.
Bottom: Cosmic-ray intensity  for test region positions located at different distances from the center of the starburst. }
\label{antipo}
\end{figure}

\begin{figure}
\center
\includegraphics[width=8.0cm]{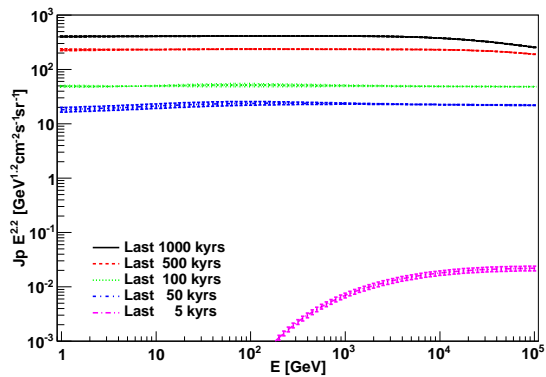}
\caption{Cosmic-ray intensity in the starburst region, at 50 pc from the center, 
and at different times, from 5000 (bottom curve) to $10^{7}$ years. The curve for $10^{7}$ years overlaps the one for $10^{6}$ years.
Errors are derived as the
standard deviation for each case after hundreds of MC realizations. }
\label{time1}
\end{figure}

Our results verify the approximate isotropy of the cosmic-ray distribution. Unless we are interested in an analysis test point that is too close to a recent accelerator location, we find that the cosmic-ray distribution is homogeneous. For example, in the top panel of Figure  \ref{antipo} we show an example of the results for antipodal positions in azimuth and zenith angle, at a distance of 100 pc from the center of the starburst. Our results also show that the cosmic-ray intensity is a function of starburst-center distance, being smaller the farther away from it we locate the test region. This is the result of a finite-size starburst sphere, and appears as a reflection of the fact that test regions closer to the limit of the star-formation environment lack the contributions of individual accelerators located beyond this distance. This is shown in the bottom panel of Figure  \ref{antipo}, for test positions located at different distances from the center of the starburst. In configurations such as this, namely a starburst sphere with a limited size, the difference between the cosmic-ray spectrum at the center and its boundary can reach up to a factor 2. This is usually neglected in the models published in the literature (see the introduction). 

We compute the cosmic-ray intensity also as a function of time from the beginning of the burst, and an example is shown in Figure \ref{time1}. There we show the cosmic-ray intensity in the starburst region at 50 pc from the center 
at different times, from 5000 (bottom curve) to $10^{7}$ years. We note that as long as the starburst is ongoing, the cosmic-ray intensity increases at all energies. Although the rate of increase of the cosmic-ray intensity slows down after a few hundred thousand years, 
it only attains steady-state after a few million years. For $\sim 10^7$ years, the cosmic ray intensity in Figure  \ref{time1}  would overlap with
 the curve corresponding to $10^{6}$ years.
It is then important to know the age of the starburst when detailed models are constructed, at least for ages up to $\tau_{pp} \sim 10^6$ years.
The latter actually depends on how dense the starburst is, in other words, on how the age of the starburst compares with $\tau_{pp}$.

\begin{figure}
\center
\includegraphics[width=8cm]{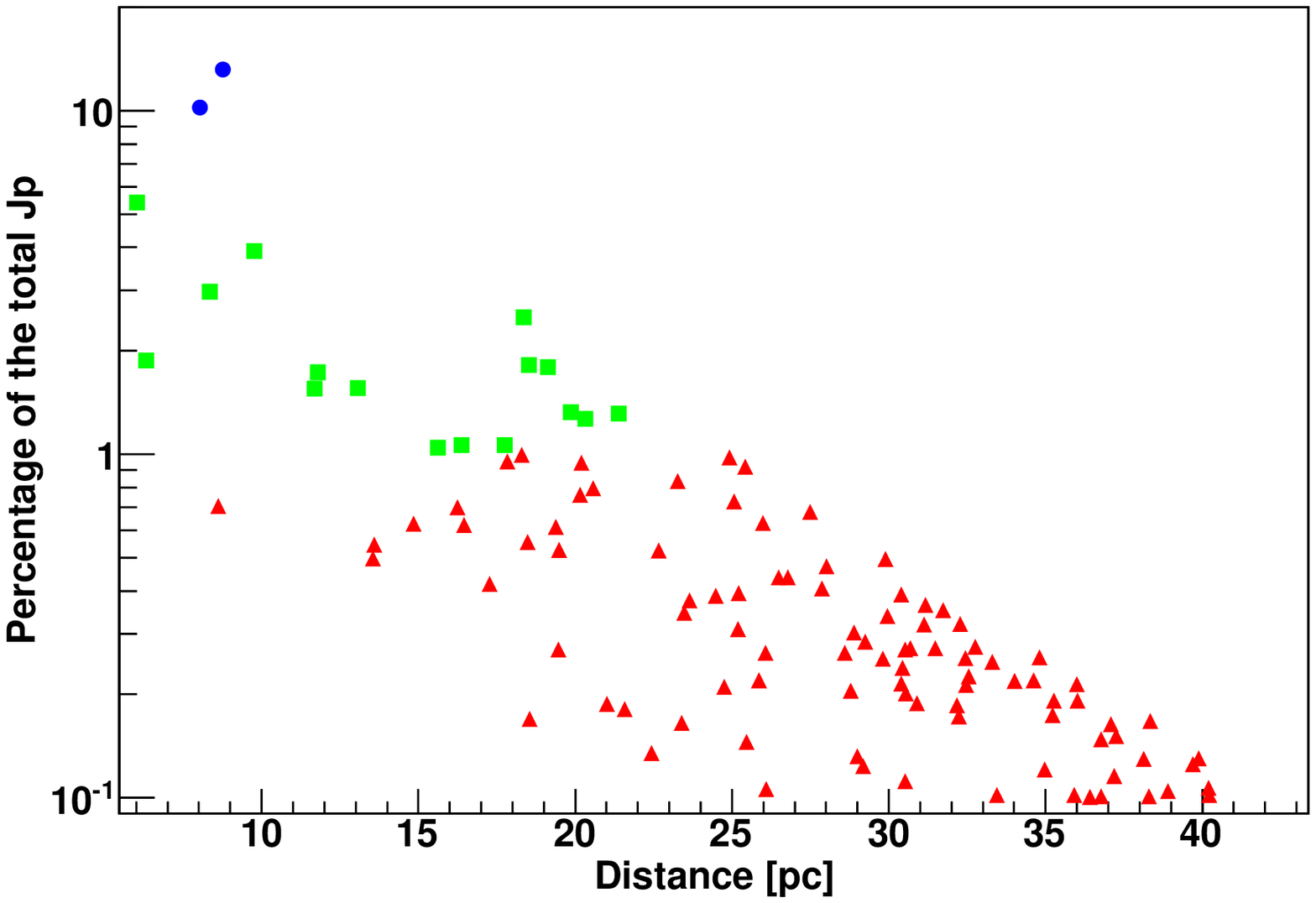}
\includegraphics[width=8cm]{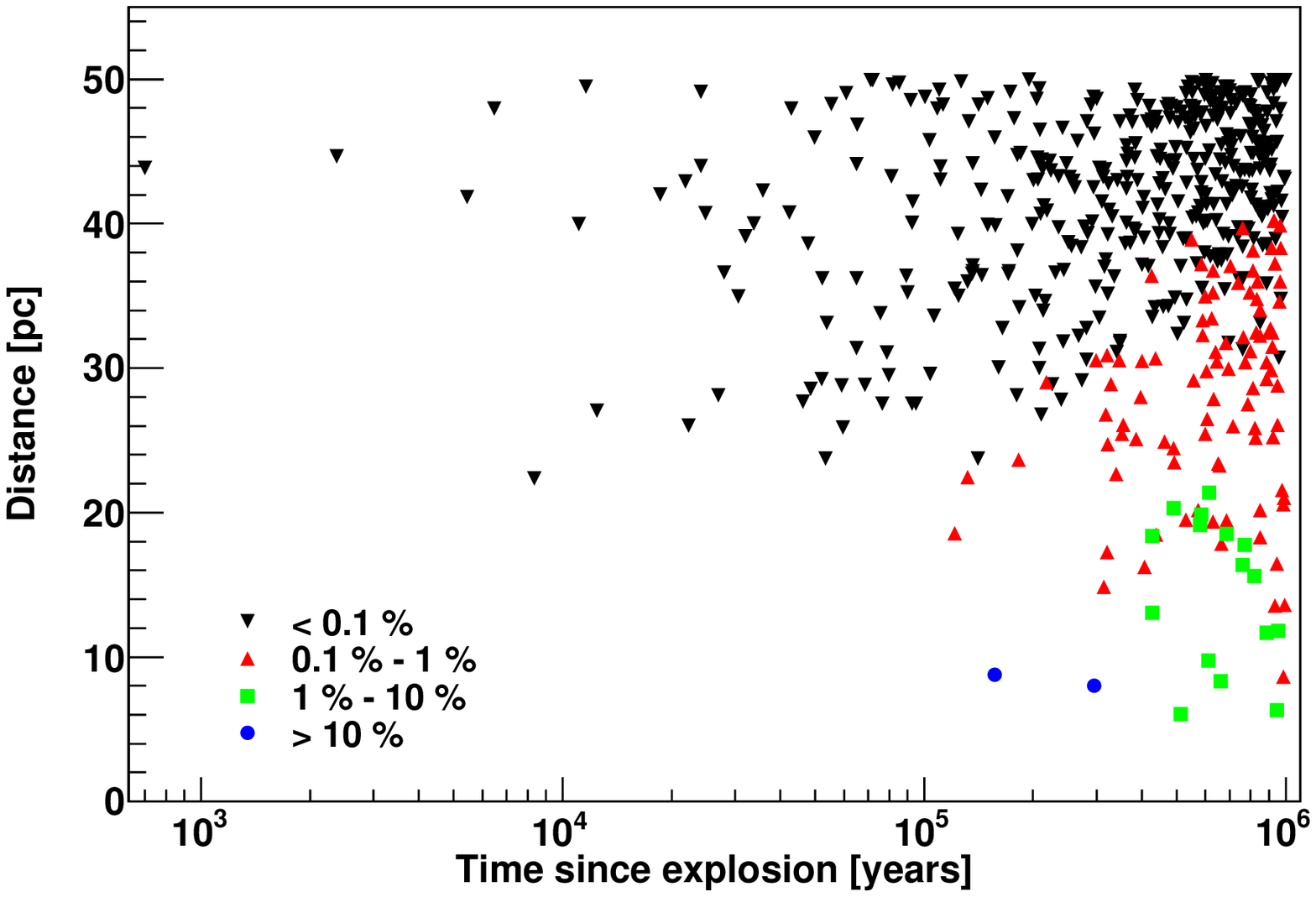}
\caption{Top: Example of cosmic-ray intensity contributions as a function of distance (around 50 pc from the test position, located at 100 pc from the center of the starburst). The blue points show accelerators that contribute more than 10\% of the total flux, in green, those contributing between 1 and 10\%, and in red those contributing between 0.1 and 1\%. Bottom: Contributions of individual accelerators surrounding a test-position at 100 pc from the center; black dots represent contributions of less than 0.1 \%.}
\label{dist-e}
\end{figure}

In our current setting and for our 
assumed explosion rate, we have about $\sim 500$ accelerators exploding over the starburst age at positions within 50 pc of any location of interest (except of course those positions located at the boundary of the starburst). These accelerators provide more than 97\%  of the cosmic ray intensity at each point. This can be seen in the example of Figure \ref{dist-e}, where we plot
the percentage of the total (integral) cosmic-ray intensity contributed by individual accelerators as a function of their distance to the test point. The contributions of individual accelerators surrounding a test position at 100 pc from the center are shown
in the bottom panel of Figure \ref{dist-e}.


\subsection{Varying the diffusion coefficient}

\begin{figure}
\center
\includegraphics[width=8cm]{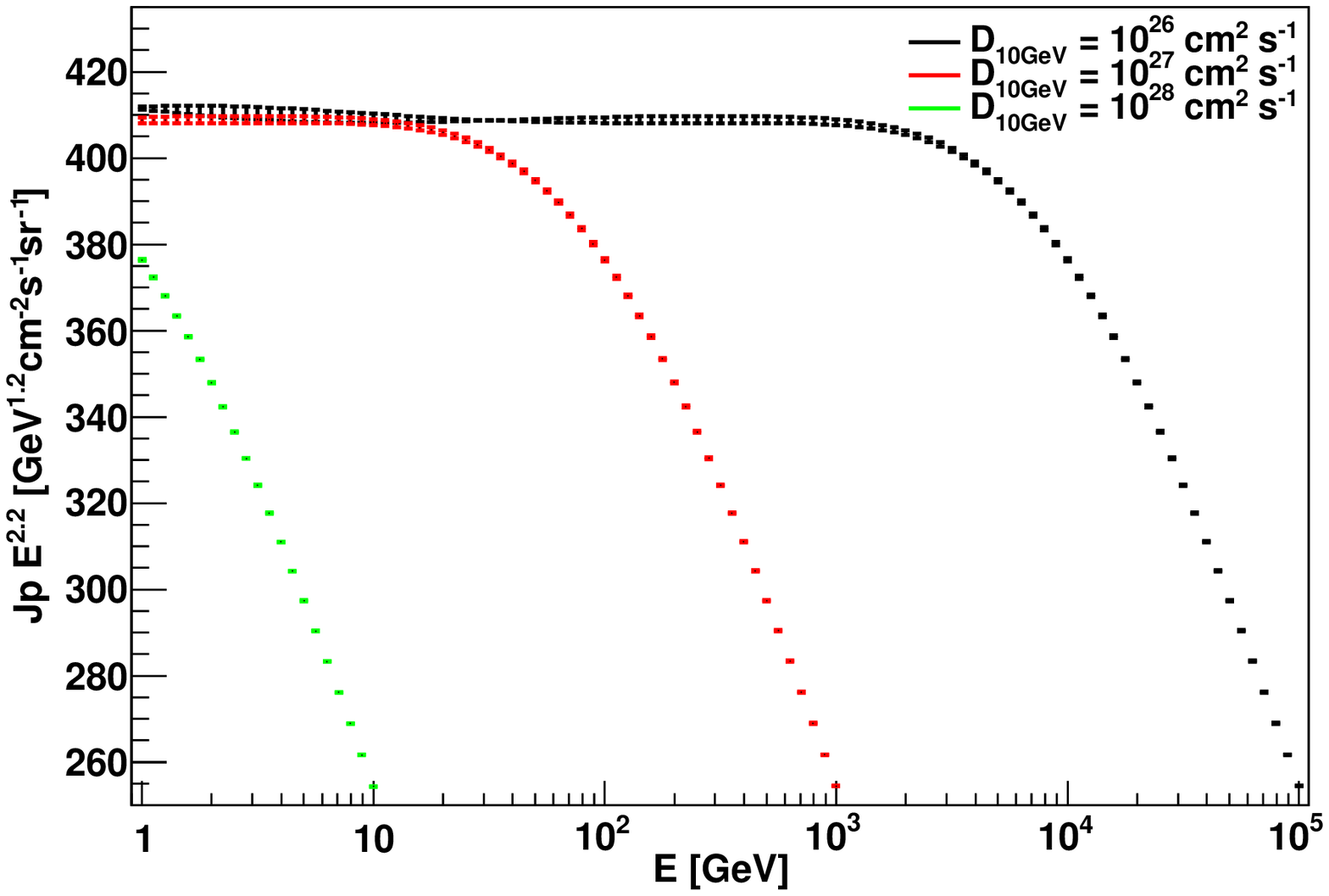}
\includegraphics[width=8cm]{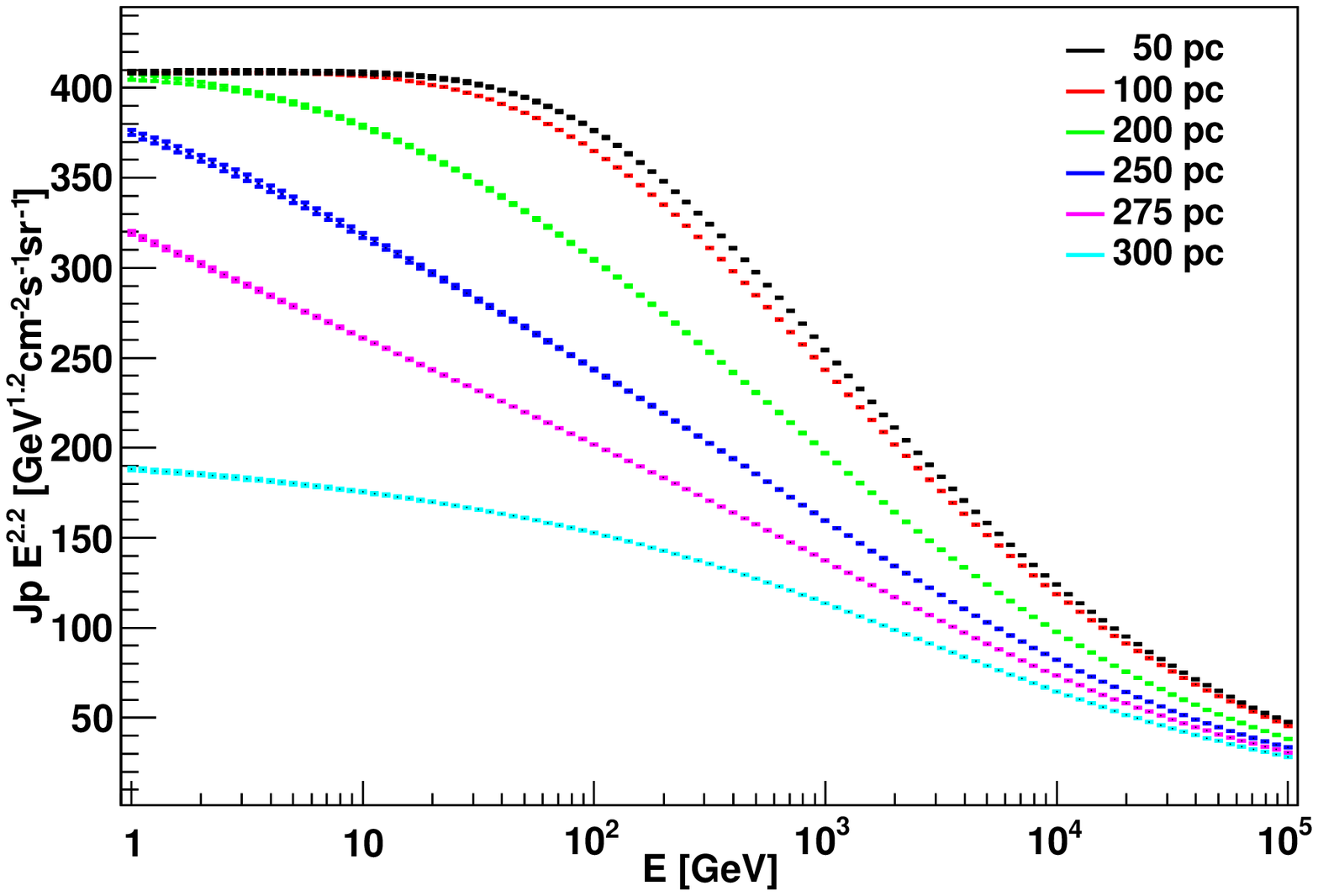}
\caption{Top: Example of cosmic-ray intensity at 50 pc from the center as a function of $D$. Bottom: Cosmic-ray intensity  for test positions located at different distances from the center of the starburst, as in Figure \ref{antipo} but for $D_{10 {\rm GeV}}=10^{27}$ cm$^{2}$ s$^{-1}$.}
\label{DD}
\end{figure}

Since the diffusion coefficient is actually unknown, we have also considered the effect of a faster diffusion (a larger $D_{10 {\rm GeV}}$, leading to a shorter timescale for diffusion of particles).
As expected, the larger $D$ is, the smaller is the cosmic-ray population at higher energies. This is exemplified at a fixed distance (50 pc) from the center of the starburst in Figure \ref{DD}. We also plot in the bottom panel of that Figure the cosmic-ray intensity  for test positions located at different distances from the center of the starburst, as in Figure \ref{antipo} but for $D_{10 {\rm GeV}}=10^{27}$ cm$^{2}$ s$^{-1}$.
These figures lead to the following conclusion: It is interesting to note that the discovery of a hard spectrum up to the highest energies would favor a small diffusion coefficient in starburst galaxies. 
If one considers that gamma-rays are $\sim $10 times less energetic than the protons that generated them, differences in the gamma-ray spectrum at say, 1--10 TeV should be clear despite possible degeneracies introduced by differences in modeling the injection or the size of the emitting region. Generally, the flatter the spectrum is at the highest energies, the better is the case for a small $D$. However, it is difficult to determine such changes from analysis of current measurements. Figure \ref{DD} shows that for $D_{10\ {\rm GeV}} = 10^{27} {\rm cm}^2\ {\rm s}^{-1}$, the $E^{2.2} J_p$ flux drops by a factor of $4$ over an energy range of 1000, between $E_p =$ 10 GeV and 10 TeV (roughly corresponding to $E_{\gamma} \sim $ 1 GeV and 1 TeV).
 That translates to a spectral steepening of 
0.2. 
This measurement is particularly suitable for the forthcoming Cherenkov Telescope Array (Actis et al. 2011), which should provide a detailed gamma-ray spectrum of nearby starburst galaxies from a few tens of GeV up to 100 TeV.

\subsection{Stability of results}

\begin{figure}
\center
\includegraphics[width=8.0cm]{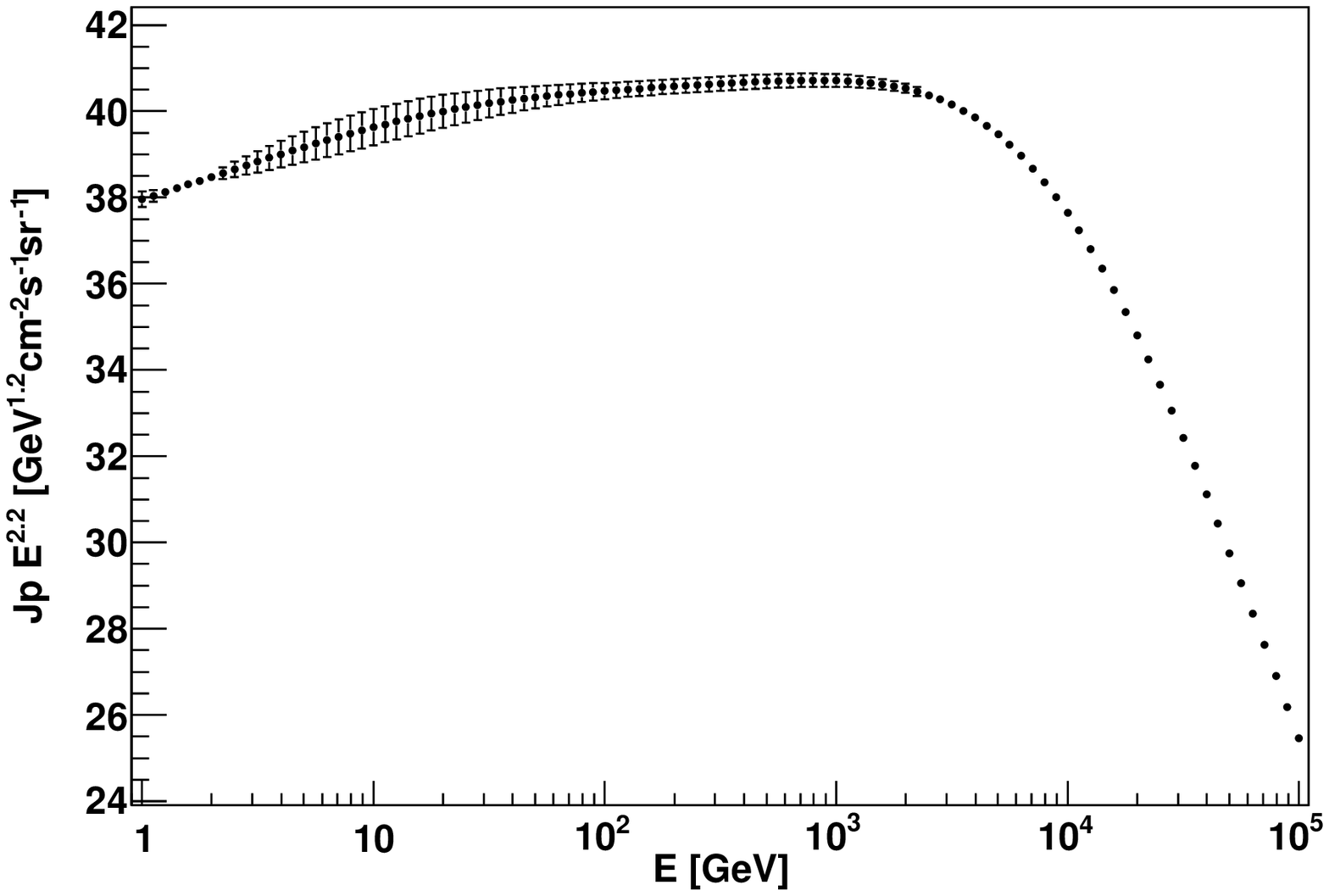}
\includegraphics[width=8.0cm]{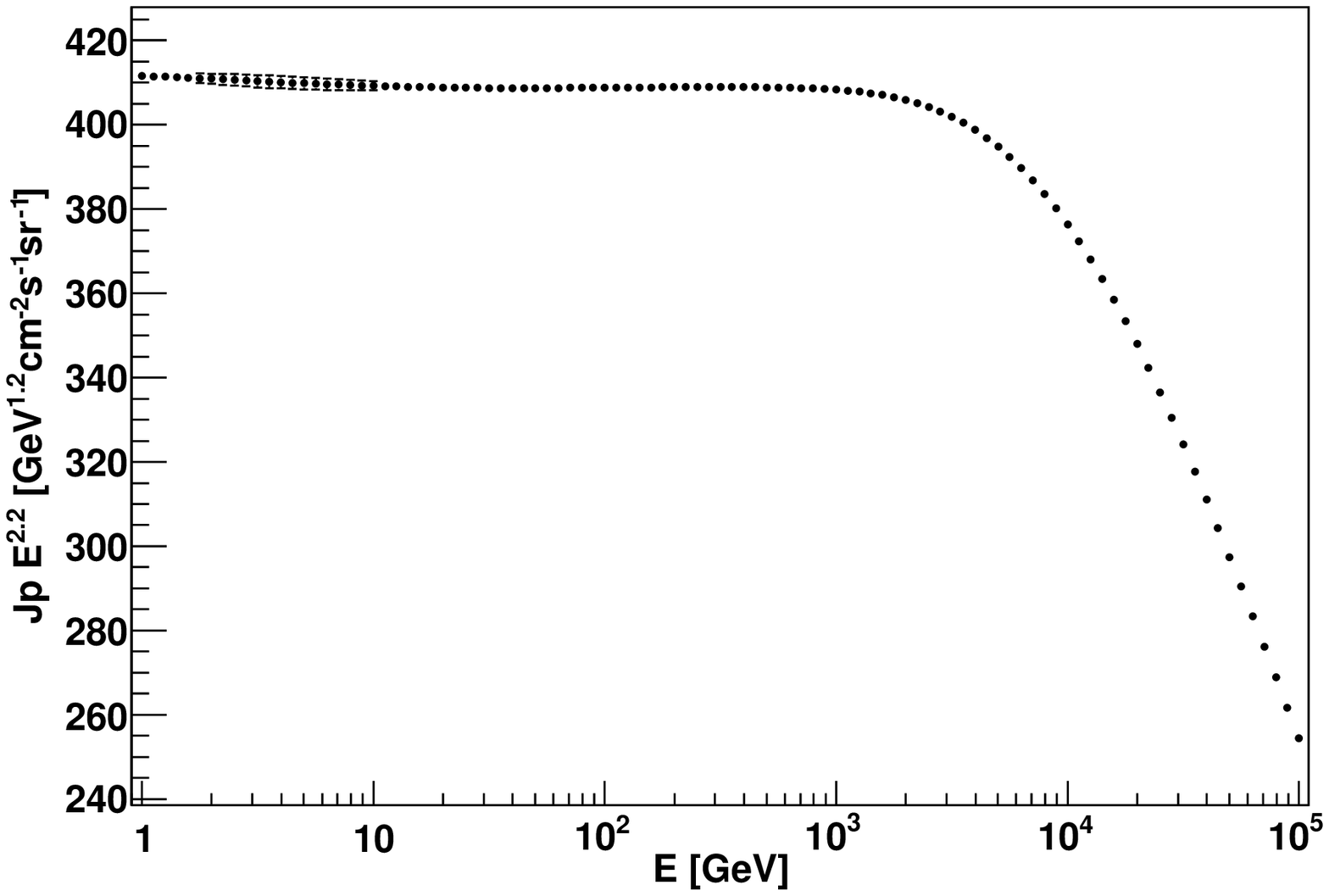}
\caption{
Cosmic-ray intensity at 50 pc from the starburst center for an accelerator appearance rate of 0.01 (top) and 0.1 (bottom) and derived standard deviation after thousands of MC realizations.}
\label{random_seed}
\end{figure}

We have tested many realizations of the set of positions where the injectors are located along the lifetime of the starburst, and the only visible difference is set by the proximity to the test point of the nearest injector. This produces differences especially at low energies, given the results commented in the previous section, but not an appreciable difference at energies beyond 100 GeV. 
At high energies, since the diffusion scales grows, the contribution is about the same irrespective of the distribution (high-energy protons diffuse significantly away from accelerators). As an example of the stability of the results obtained out of a given realization we plot the cosmic-ray intensity
at 50 pc from the starburst center for an accelerator appearance rate of 0.01 and 0.1 (see Figure \ref{random_seed}). The errors  are the
derived standard deviation after thousands of MC realizations. Points at higher energies show barely any error. 

%

\begin{figure}
\center
\includegraphics[width=8.0cm]{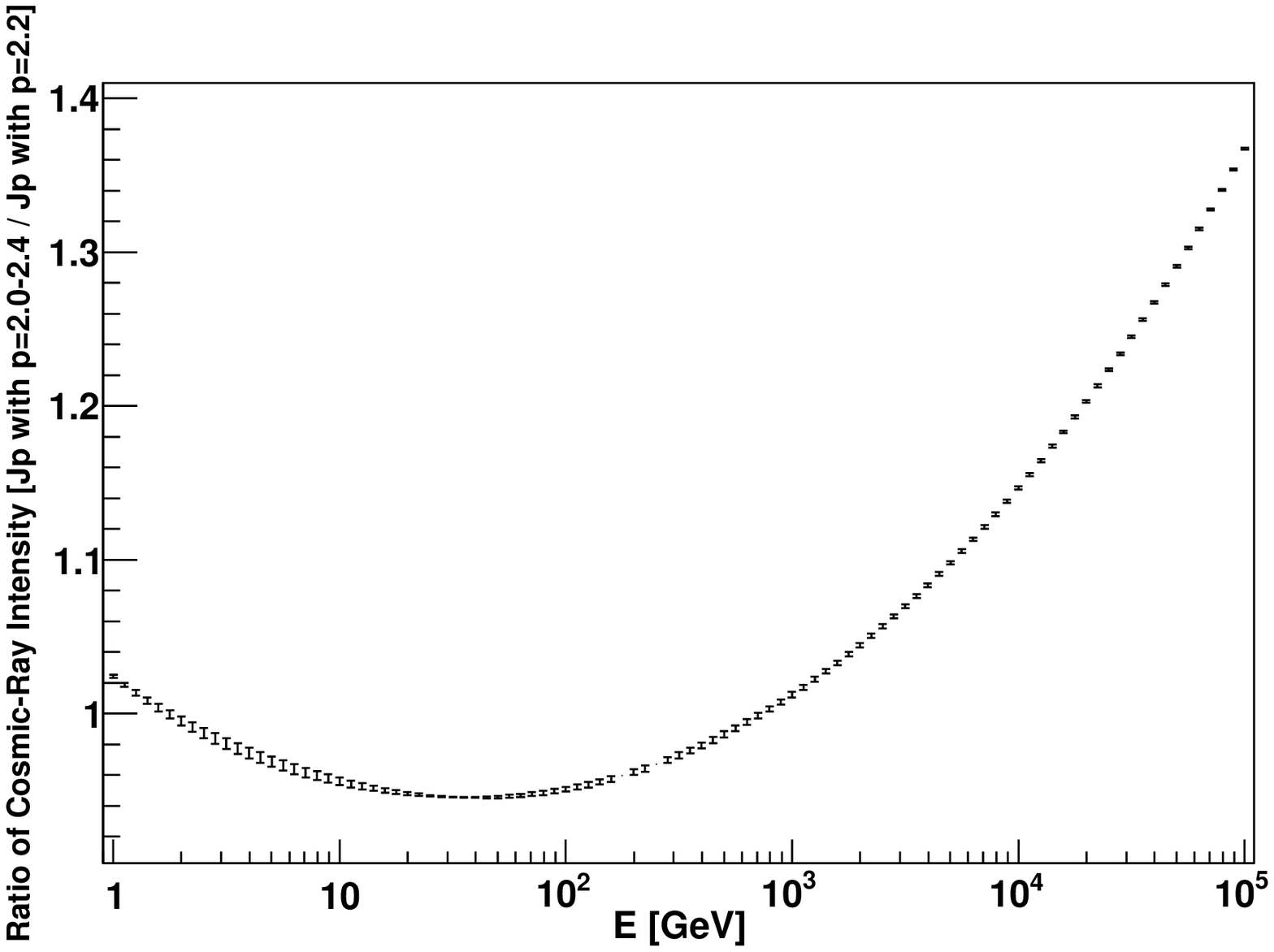}
\includegraphics[width=8.0cm]{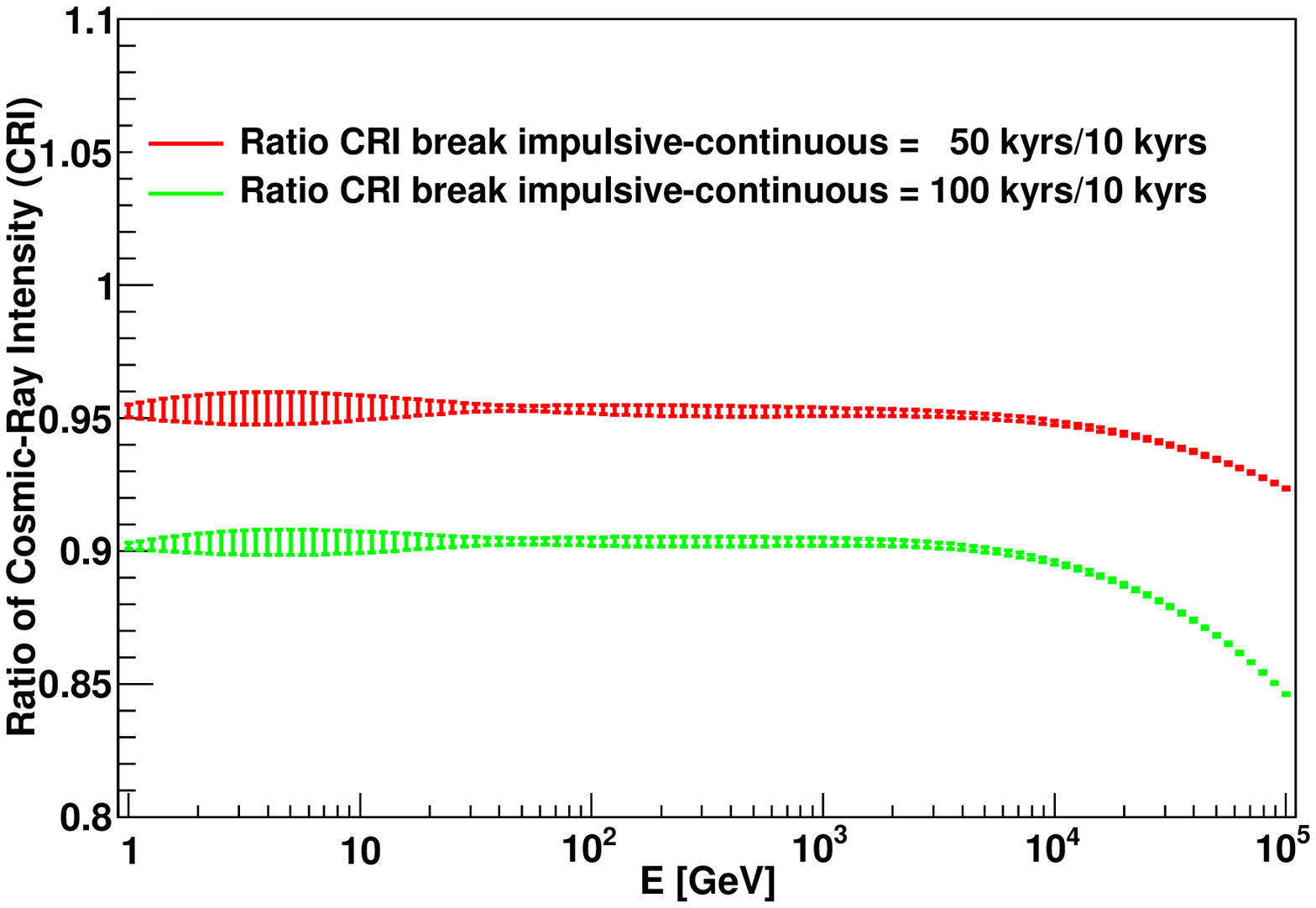}
\caption{
Ratio of cosmic-ray intensity for configurations allowing a variable [2.0-2.4] and fixed [2.2] slope of the injection spectra, and its derived standard deviation after thousands of MC realizations of the accelerators positions and times of their appearance.
Botton:
Ratio for cosmic-ray intensity with different regime breaks.
In both cases the test point was chosen at 50 pc from the center of the starburst, as an example.  }
\label{ratio_alpha_random_fix}
\end{figure}

We ran two additional tests. The first one is on the influence of a fixed value of the slope of the cosmic-ray injection for all accelerators instead of a variable one, which we have 
chosen in the range 2.0 -- 2.4 { (see, e.g. Bell et al. 2011)}. That is, in the second case each injector has a different (randomly assigned) slope on its cosmic-ray spectral injection. The results are depicted in Figure \ref{ratio_alpha_random_fix}
where we see that the change is visible especially at high energies, where it can reach up to $\sim$ 40 \% due to the influence of harder injectors.

We have also tested the influence of a different cut between the impulsive and continuous assumptions for the accelerators (i.e., a different regime break), by studying the stability of the results for breaks at 50000 and 100000 years. The second panel of Figure \ref{ratio_alpha_random_fix}
shows the ratio for the cosmic-ray intensity with different regime breaks with that considered earlier
at a test point 50 pc away from the starburst center. The larger the regime break, the more accelerators are considered continuous injectors. This explains why the cosmic-ray intensity is smaller the larger the break point:
the higher the energy, the shorter is the time a particle has
to reach the test point in the case of a continuous accelerator,
producing a difference 
of up to a few percent in the case of a 50000 years break.

\subsection{Terminated star-formation burst}

\begin{figure}
\center
\includegraphics[width=8.0cm]{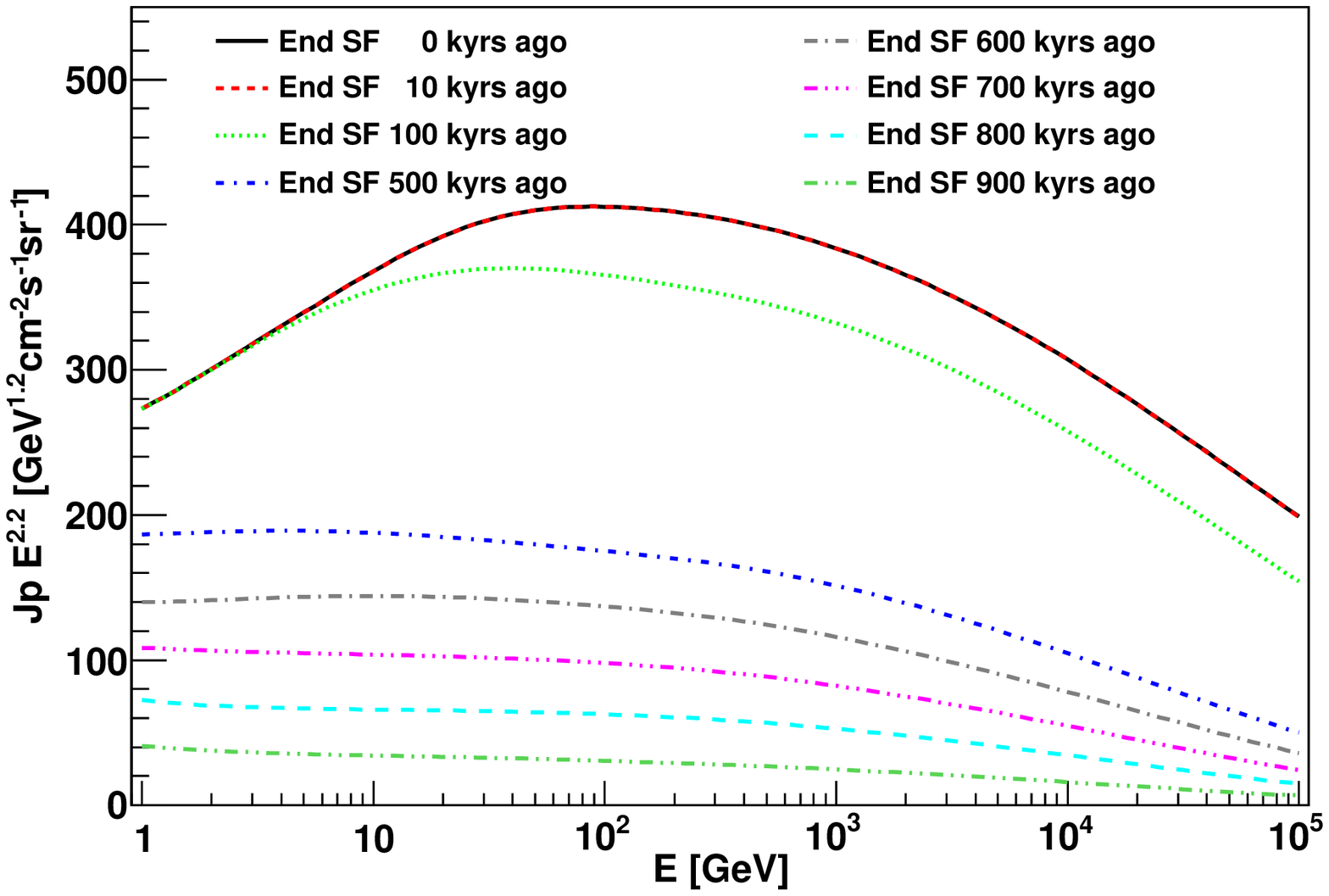}
\includegraphics[width=8.0cm]{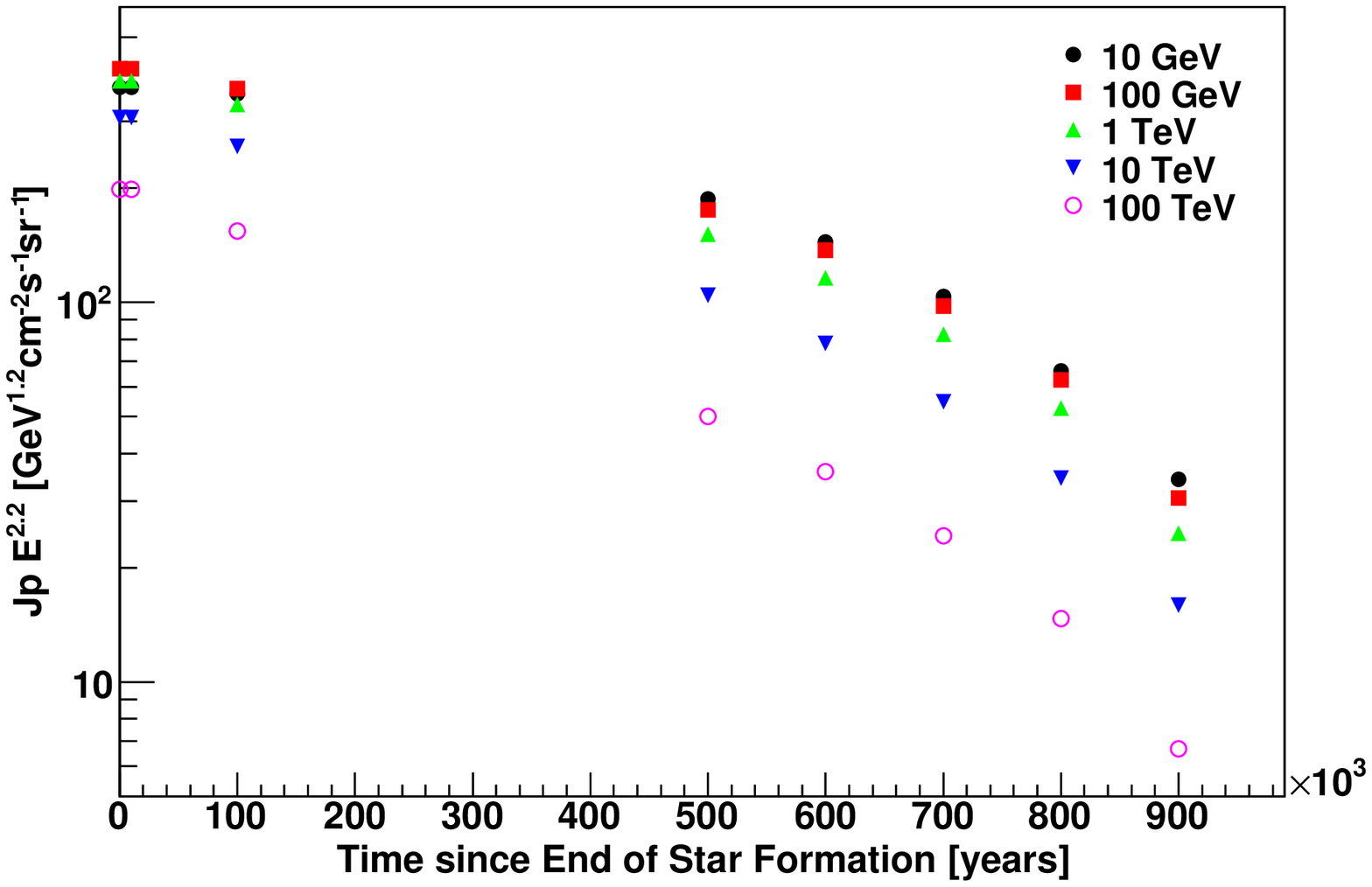}
\caption{Top: Cosmic-ray spectrum today for different times at which the star formation is exhausted. Bottom: decay of the cosmic-ray intensity as a function of time and energy.  
 }
\label{ex1}
\end{figure}

To exemplify the cosmic-ray decay, we analyze the situation in which the star formation is exhausted in the whole starburst.
Figure \ref{ex1} shows how fast the cosmic-ray spectrum dies out when the star formation stops as a function of time from the
end of the burst. It is interesting to note that the cosmic-ray spectrum, even in cases in which the last accelerator appears $5\times 10^5$ years ago, is only a factor of 2 less than what it would be under a continuous star forming process. 
The timescale for decay in this model is of order $\sim \tau_{pp}$, as might be expected if pion production is the only loss.
The decay of the cosmic ray spectrum is slow
as shown in the bottom panel of Figure \ref{ex1}. We obtained similar results also for the case in which there is a remaining, residual star formation only in the central part of the original region.

\section{The case of NGC 253}

We now consider a core-enhanced star-formation rate distribution where the starburst geometry is divided into a central sphere (or innermost starburst), in which the star formation is maximal, and a surrounding disk. Such a situation is typical in starbursts galaxies, and has been considered, for instance, for NGC 253, where 
infrared, millimeter, and centimeter observations show that the 
central region dominates ongoing star formation (e.g., Ulvestad \& Antonucci 1997; Ulvestad 2000). 
Current estimates of the gas mass in the central 20--50 arcsec
region range from $2.5 \times 10^7 M_\odot$ (Harrison et al.
1999) to $4.8 \times 10^8 M_\odot$ (Houghton et al. 1997); see Bradford
et al. (2003), Sorai et al. (2000), and Engelbracht et al. (1998). 
We also note that a detailed numerical treatment of the steady-state distribution of cosmic ray electrons and protons in NGC 253 was recently carried out by  Paglione et al. (1996), Domingo-Santamaria \& Torres (2005), and Rephaeli, Arieli, \& Persic (2010). { In the latter case, their modified GALPROP code follows the evolution of particle energy and spatial distribution functions in the context of a diffusion-convection model. No time-dependent analysis of the contribution of individual accelerators is included.}

Here, following 
Domingo-Santamar\'ia \& Torres (2005)
we will assume, 
in agreement with the mentioned measurements, 
that within the central 200 pc (100 pc radius), a disk of 70 pc
hosts $3 \times  10^7 M_\odot$ of molecular mass, uniformly distributed, with a density of $\sim 600$ cm$^{-3}$. In this region, the supernova explosion rate is taken as 0.08 year$^{-1}$. We have also considered a lower 
value for the accelerator appearance in the core (0.02 yr$^{-1}$). This is motivated by the relationship between supernova rate and IR luminosity (Lacki et al. 2011) and also by the fact that Melo et al. (2002) indicate that 50\% of the IR luminosity (and 70\% of the star-formation) is in the starburst and 30\% of the IR and star-formation is in the surrounding disk. Williams \& Bower (2010) also find that about half of the GHz radio emission is extended (from the disk) and half comes from the starburst core, which,  to the extent the FIR-radio correlation holds,  suggests a roughly equal star-formation rates in the starburst and the surrounding disk.

Additional mass with an average density of $\sim $ 50 cm$^{-3}$ is assumed to be in the central kpc outside the innermost region, but subject to a smaller injector rate of 0.01 year$^{-1}$, or $\sim$ 10\% of that found in the most powerful nucleus (Ulvestad 2000). We assume that each injector accelerates cosmic-rays with a spectral slope between 2.2 and 2.4, randomly assigned.

\begin{figure}
\center
\includegraphics[width=8.0cm]{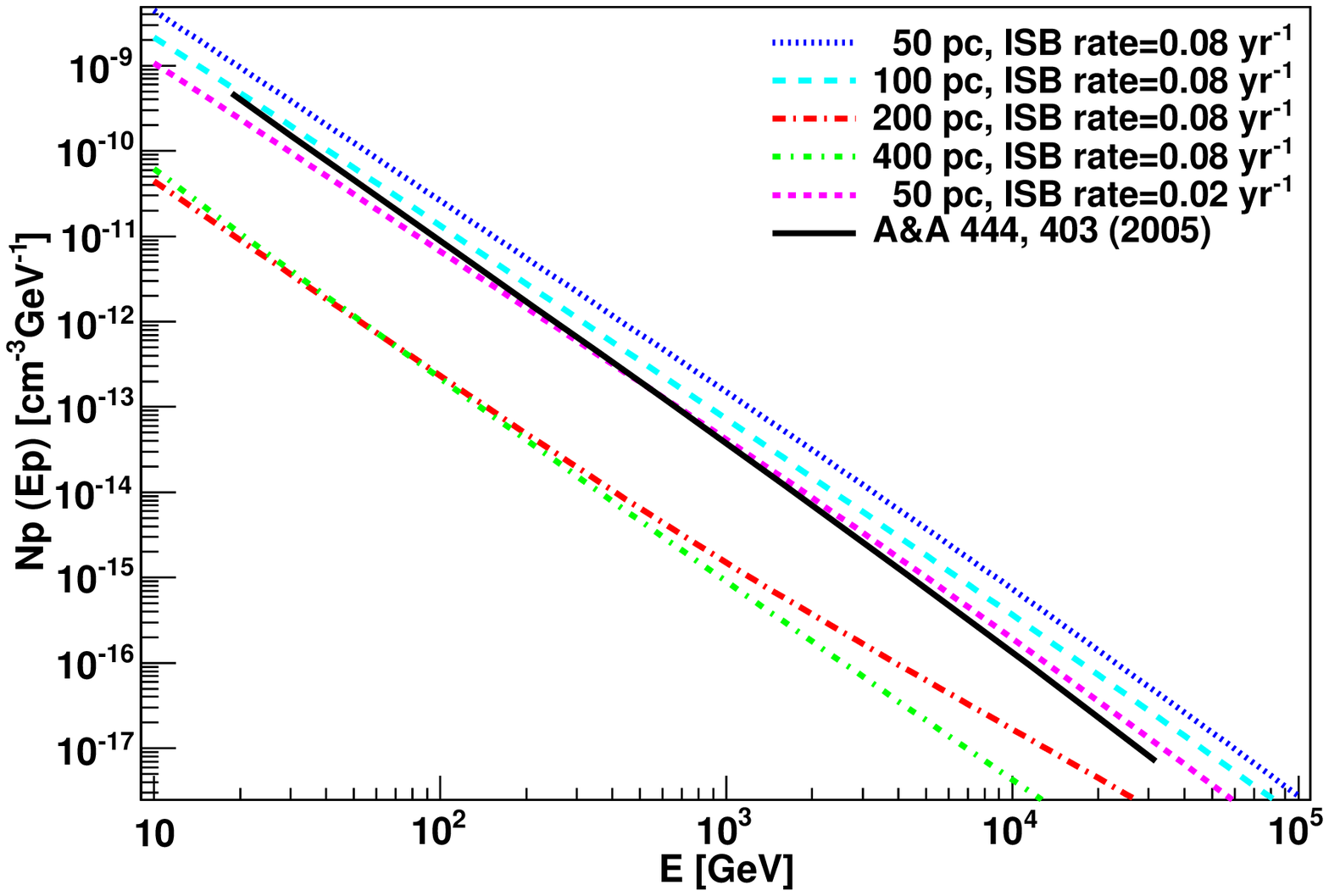}
\includegraphics[width=8.0cm]{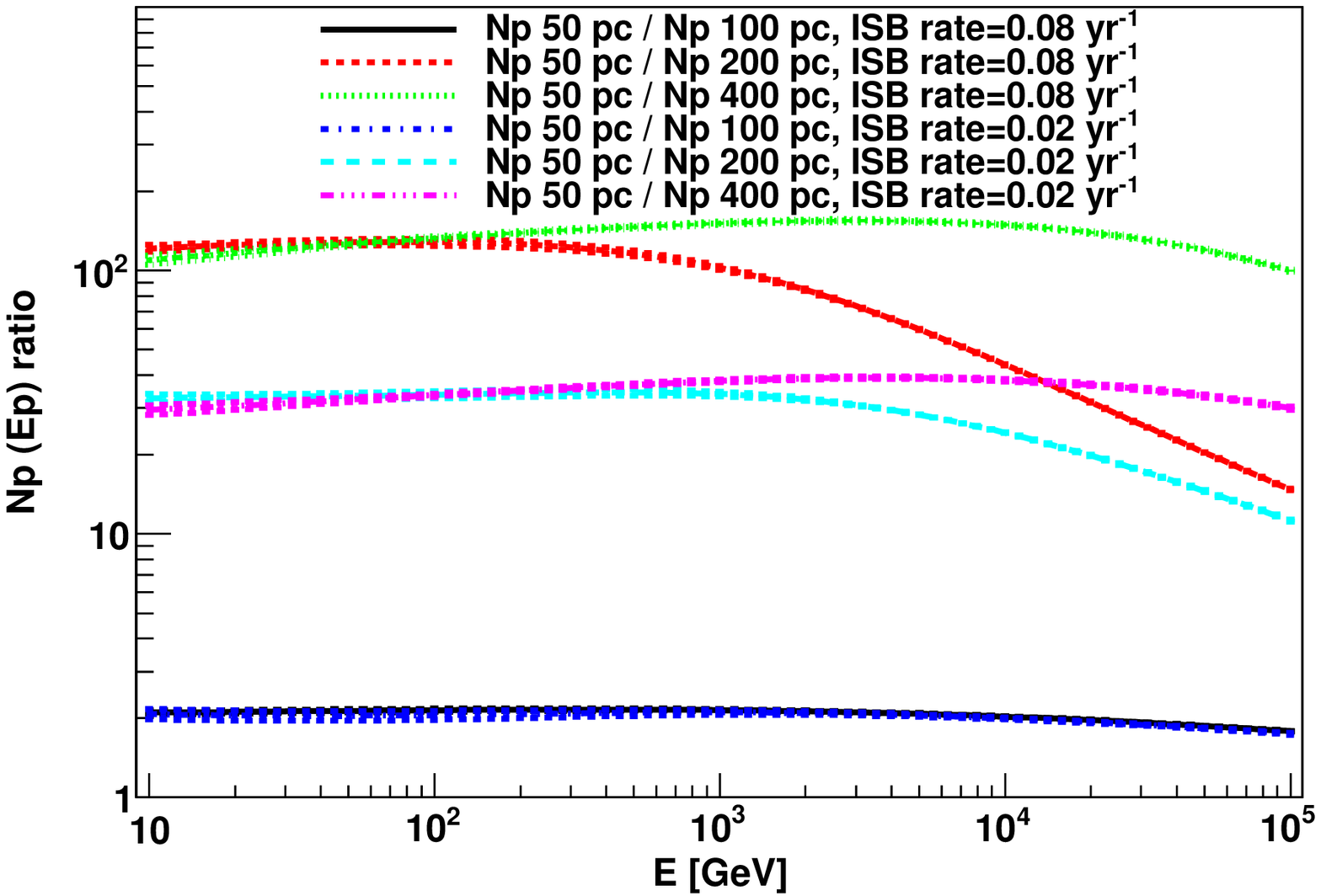}
\caption{Model for cosmic-ray distribution in NGC 253. Shown in the upper panel are proton spectra at high energies,
and at different distances from the galaxy center, along the plane. The bottom panel shows the ratio between the cosmic-ray distribution at 50 pc in the ISB and the one at different distances, within the central core and along the disk, as a function of energy. Two different accelerator appearance rates are considered. 
 }
\label{ngc}
\end{figure}

Figure \ref{ngc} shows the result for the cosmic-ray distribution at high energies, and at different distances from the center of NGC 253, 
within and outside the innermost starburst region
 with the highest accelerator appearance rate. Results for different accelerators appearance rate in the innermost starburst are shown. Each value of $N_p$ is obtained by summing up all contributions coming from accelerators located in the innermost starburst as well as those located in the disc.
 The top panel also shows the average cosmic-ray distribution obtained by Domingo-Santamar\'ia \& Torres (2005), based on the treatment described in the introduction. 
The impact of a core-enhanced 
rate distribution is evident, with a significant change, of more than an order of magnitude at 1 TeV, in the cosmic-ray density when the distance is increased from 50 pc to 200 pc from the center (i.e., from within to outside the innermost nucleus). There is in addition a significant change in the cosmic-ray distribution further outside of the nucleus, the plot shows it for distances from 200 to 400 pc along the surrounding disc. 
The bottom panel of Figure \ref{ngc} shows the ratio between the cosmic-ray distribution in the sphere and the one at different distances from the center as a function of energy. It can seen be again that large values of the ratio are obtained; even for a smaller difference between the rates of the (innermost starburst) ISB and that of the surrounding disc. Thus, one-zone models are a rough approximations to this situation.

\section{Concluding remarks}

The results presented in this paper are based
on a step-by-step approach for the computation of the build-up of the cosmic-ray spectrum in star forming 
environments: The respective contribution of each accelerator injecting cosmic rays over the history of the star forming region is individually accounted for.
The total spectrum is then obtained by summing up all individual contributions, taking into account an energy, time, and space-dependent propagation of the cosmic-rays in the starburst environment. Even with the caveat of not yet considering outflows, 
the total spectrum shows two features that are found from gamma-ray observations of starburst galaxies, 1) that the average photon index of the current cosmic-ray spectrum in starbursts 
is  hard, even though the cosmic ray spectrum of the individual injectors suffer significant steepening due to their propagation, and 2) put in context the fact that the integrated cosmic-ray spectrum (and its gamma-ray yield) scale linearly with the star formation rate. The latter can also be considered an observed fact (see Abdo et al. 2011) given that the gamma-ray luminosity of nearby star forming galaxies seem to scale linearly with the radio continuum and infrared luminosity, both established tracers of star formation. Our simulations shows that
the discovery of a hard spectrum up to the highest energies would favor a small diffusion coefficient. 
Our results also prove that in the calorimetric limit (with no cosmic-ray advection), homogeneity is reached (typically within 20\%) across the whole starburst region. However, 
values of center-to-edge intensity ratios can amount to a factor of several. Differences between the common homogeneous assumption for the cosmic-ray distribution and our models are larger in the case of two-zone geometries, such as a central nucleus with a surrounding disc.
We have also found that the decay of the cosmic-ray density after the starburst process is terminated is slow, and even approximately 1 Myr after the burst ends (for a gas density of 35 cm$^{-3}$) it may still be within an order of magnitude of its peak value.

\subsection*{Acknowledgments}

This work was supported by the  grants AYA2009-07391 and SGR2009-811, as well as the Formosa program
TW2010005 and iLINK program 2011-0303. AC is a
member of the Carrera del Investigador Cient\'ifico of CONICET, Argentina. BCL is supported by a Jansky fellowship from the NRAO.  NRAO is operated by Associated Universities, Inc., under cooperative agreement with the National Science Foundation.
We thank Ana Y. Rodriguez-Marrero, A. Caliandro, G. Pedaletti, \& O. Reimer  for discussions.

\label{lastpage}

\begin{thebibliography}{}

\bibitem{} Abdo A., 2010, ApJ Letters 709, 152
\bibitem{} Abdo A., 2011, ApJ submitted 

\bibitem{} Acciari V. A., et al. 2009, Nature, 462, 770 1371
\bibitem{} Acero F., et al. 2009, Science, 326, 1080
\bibitem{}Actis M. et al. 2011, Exp. Astronomy 32, 193

\bibitem{} Aharonian F. A., \& Atoyan A. M. 1996, A\&A 309, 91


\bibitem{} { Bell, A. R. 1978, MNRAS, 182, 443}
\bibitem{} { Bell, A. R.; Schure, K. M.; \& Reville, B. 2011, MNRAS, 418, 1208 }

\bibitem{} { Berezhko, E.~G., \& Ellison, D.~C.\ 1999, ApJ, 526, 385}

\bibitem{} Blom, J. J., Paglione, T. A., \& Carrami–ana, A. 1999, 516, 744
\bibitem{} Bradford, C. M., Nikola, T., Stacey, G. J., et al. 2003, ApJ, 586, 891

\bibitem{} de Cea del Pozo E., Torres D. F., \& Rodriguez-Marrero A. Y., 2009, ApJ 698, 1054
\bibitem{} Domingo-Santamar\'ia, E., \& Torres, D. F. 2005, A\&A, 444, 403


\bibitem{} { Ellison, D.~C., Berezhko, E.~G., \& Baring, M.~G.\ 2000, ApJ, 540, 292}
\bibitem{} { Ellison, D. C., Decourchelle, A., \& Ballet, J. 2004, A\&A, 413, 189}
\bibitem{} Engelbracht C. W., Rieke M. J., Rieke G. H., Kelly, D. M., \&
Achterman, J. M. 1998, ApJ, 505, 639


\bibitem{} { Fujita, Y., Ohira, Y.,
Tanaka, S.~J., \& Takahara, F.\ 2009, ApJ Letters 707, L179}
\bibitem{} { Fujita, Y., Ohira, Y., \& Takahara, F.\ 2010, ApJ Letters, 712, L153}

\bibitem{} Gabici S. 2010, In proceedings of ICATPP Conference on Cosmic Rays for Particle and Astroparticle Physics, Villa Olmo, Como 7-8 October 2010, arXiv:1011.2029

\bibitem{}  { Gaisser T. K. 1990, Cosmic Rays and Particle Physics, Cambridge University Press, Cambridge, UK }
\bibitem{} Ginzburg V. L. \& Syearovatskii S. I. 1964, ``The origin of
cosmic rays", Pergamon Press, Oxford, England.

\bibitem{} Harrison, A., Henkel, C., \& Russel, A. 1999, MNRAS, 303, 157
\bibitem{} Houghton, S., Whiteoak, J. B., Koribalski, B., et al. 1997, A\&A, 325,
923

	
\bibitem{} 	Lacki B. C., Thompson T. A. \& Quataert, E. 2010, ApJ 717, 1
\bibitem[Lacki et al.(2011)]{Lacki11} Lacki, B.~C., Thompson, T.~A., Quataert, E., Loeb, A., \& Waxman, E.\ 2011, ApJ 734, 107
\bibitem[Loeb \& Waxman(2006)]{Loeb06} Loeb, A. \& Waxman, E. 2006, Journal of Cosmology and Astroparticle Physics 5, 3


\bibitem[Makiya et al.(2011)]{Makiya11} Makiya, R., Totani, T., \& Kobayashi, M.~A.~R.\ 2011, ApJ 728, 158
\bibitem{} Melo, V. P., Perez-Garcia, A. M., Acosta-Pulido, J. A., Mu\~noz-Tu\~non,
\& Rodriguez Espinosa, J. M. 2002, ApJ, 574, 709
	

\bibitem{} Ormes J. F., Ozel M. E., \& Morris D. J., 1988, ApJ, 334, 722


\bibitem[Papadopoulos(2010)]{Papadopoulos10} Papadopoulos, P.~P.\ 2010, ApJ 720, 226
\bibitem{} Paglione T. A. D., Marscher A. P., Jackson J. M. \&
Bertsch D. L. 1996, ApJ 460, 295

\bibitem{} Persic, M., Rephaeli, Y., \& Arieli, Y. 2008, A\&A, 486, 143
\bibitem{} Persic, M., Rephaeli, Y., 2010 MNRAS 403, 1569
\bibitem{} { Ptuskin, V.~S., \& Zirakashvili, V.~N.\ 2005, A\&A, 429, 755}



\bibitem{} Rephaeli Y., Arieli, A. \& Persic. M., 2010, MNRAS, 401, 473

\bibitem[Socrates et al.(2008)]{Socrates08} Socrates, A., Davis, S.~W., \& Ramirez-Ruiz, E.\ 2008, ApJ  687, 202
\bibitem{} Sorai, K., Nakai, N., Nishiyama, K., \& Hasegawa, T. 2000, Publ.
Astron. Soc. Japan, 52, 785
\bibitem{} Stecker, F. W. 1971, Cosmic Gamma Rays (Baltimore: Mono)
\bibitem{} { Strong A. \& Moskalenko I. 1998, ApJ 509, 212}


\bibitem[Thompson, Quataert, \& Waxman(2007)]{Thompson07} Thompson, T. A., Quataert, E., Waxman, E. 2007, ApJ 654, 219
\bibitem{} Torres, D. F. 2004, ApJ 617, 966
\bibitem{} Torres, D. F., Rodriguez-Marrero A., \& de Cea del Pozo E. 2010, MNRAS 408, 1257 


\bibitem{} Ulvestad, J. S. 2000, ApJ, 120, 278 
\bibitem{} Ulvestad, J. S., \& Antonucci, R. R. J. 1997, ApJ, 488, 621


\bibitem{} Williams, P. K. G. \& Bower, G. C. 2010, ApJ 710, 1462





\end{thebibliography}
\end{document}